\documentclass[preprint,review,12pt,authoryear]{elsarticle}

\usepackage{amssymb}
\usepackage{subfigure}
\usepackage{graphics}

\journal{Powder Technology}

\begin{document}

\begin{frontmatter}



\title{Slow relaxation behavior of cohesive powders}

\author[label1]{Olukayode I. Imole\corref{cor1}}
\ead{o.i.imole@utwente.nl}

\author[label2,label3]{Maria Paulick}
\author[label3]{Martin Morgeneyer}
\author[label1]{Vanessa Magnanimo}
\author[label4]{ Bruno E. Ch{\' a}vez Montes}
\author[label5,label6]{Marco Ramaioli}
\author[label2]{Arno Kwade}
\author[label1]{Stefan Luding}

\address[label1]{Multi Scale Mechanics (MSM), CTW, MESA+, University of Twente, P.O. Box 217, 7500 AE Enschede, The Netherlands.}
\address[label2]{Institute for Particle Technology (iPAT), TU Braunschweig, Volkmaroderstr. 4-5, 38104 Braunschweig, Germany.}
\address[label3]{Universit{\' e} de Technologie de Compi{\` e}gne, B.P. 20.529, 60200 Compi{\` e}gne, France.}
\address[label4]{Nestl{\' e} Product Technology Centre Orbe, Rte de Chavornay 3, CH-1350 Orbe, Switzerland.}
\address[label5]{Nestl{\' e} Research Center, Lausanne, Switzerland.}
\address[label6]{Department of Chemical and Process Engineering, FEPS (J2), University of Surrey, Guildford GU2 7XH, United Kingdom.}
\cortext[cor1]{Corresponding author}  

%

\begin{abstract}
We present findings from uniaxial (oedometric) compression tests on two cohesive industrially relevant granular materials (cocoa and limestone powder). Experimental results are presented for the compressibility, 
tested with two devices -- the FT4 Powder Rheometer and the custom made lambdameter. We focus on the stress response and the slow relaxation behavior of the cohesive samples tested.
After compression ends, at constant volume, the ongoing stress relaxation is found to follow a power law consistently for both cohesive powders and for the different testing equipments. 
A simple (incremental algebraic evolution) model is proposed for the stress relaxation in cohesive powders, which includes a response timescale along with a second, dimensionless relaxation parameter. 
The reported observations are useful for both the improvement of discrete element simulations and
constitutive macroscopic models for cohesive granular materials.

\end{abstract}


\end{frontmatter}


\newcommand{\fmax} { f^\mathrm{max} }
\newcommand{\smax} { \sigma^\mathrm{max} }
\newcommand{\svmax} { \sigma^\mathrm{max}_{\mathrm{v}} }
\newcommand{\pmax} { p_\mathrm{max} }
\newcommand{\dpress} { \Delta {p} }
\newcommand{\dpresseps} { \Delta {p} (\dot{\varepsilon}) }
\newcommand{\dpressrelax} { \Delta {p} (r) }
\newcommand{\dstresseps} { \Delta {\sigma_{\mathrm{v}}} (\dot{\varepsilon}) }
\newcommand{\dstressrelax} { \Delta {\sigma_{\mathrm{v}}} (r) }
\newcommand{\tr} {\tau_{R} }

\newcommand{\vstress} {\sigma_{ \mathrm{v}}}
\newcommand{\ostress} {\sigma_{ \mathrm{vo}}}
\newcommand{\vstressmax} {\sigma_{ \mathrm{v}}^{\mathrm{max}}}
\newcommand{\evol} {\varepsilon_{ \mathrm{vol}}}

\section{Introduction and Background}
\label{intro}

Granular materials are omnipresent in nature and widely used
in various industries ranging from food, pharmaceutical, agriculture 
and mining -- among others. In many granular systems
interesting phenomena like dilatancy, anisotropy, shear-band localization, 
history-dependence, jamming and yield have attracted significant 
scientific interest over the past decade 
\cite{luding2008cohesive, imole2013force, alonsomarroquin2004ratcheting}.
The bulk behavior of these materials depends on the behavior of 
their constituents (particles) interacting through contact forces. To
understand their deformation behavior,  various 
laboratory element tests can be performed 
\cite{schwedes2003review,midi2004dense}. 
Element tests are (ideally homogeneous) macroscopic tests in which one can control the stress and/or strain 
path. Such macroscopic experiments are important ingredients in developing and calibrating constitutive relations and they complement
numerical investigations of the behavior of granular materials, e.g. with the discrete element method \cite{luding2008cohesive}. 
Different element test experiments on packings of bulk solids have been realized experimentally in the biaxial box \cite {morgeneyer2003make,rock2008steady,morgeneyer2003investigation} 
while other deformations modes, namely uniaxial  and volume conserving shear have also been reported \cite {saadatfar2012mapping,yun2011evolution,philippe2011settlement}.  
Additionally, element tests with more complex, non-commercial testers have been reported in 
literature \cite{harder1985development,janssen2003measurements, head1998manual, bardet1997experimental}, 
even though their applications are restricted for example to the testing of geophysically relevant materials at relatively higher consolidating stresses.

The testing and characterization of dry, non-sticky powders is well established. For example, rotating drum experiments to determine the dynamic angle of repose have 
been studied extensively  as a means  to characterize non-cohesive powders \cite{ristow1996dynamics,baumann1996angle,cantelaube1995radial}, even though these tests 
are not well defined with respect to the powder stress and strain conditions. The main challenge comes when the powders are sticky, cohesive and less flowable 
like those relevant in the food industry.  For these powders, dynamic tests are difficult to perform due to contact adhesion and clump formation. One possibility to overcome 
this challenge is to perform confined quasi-static tests at higher consolidation stresses.

One element test which can easily be realized (experimentally and numerically) is the uniaxial (or oedometric) compression (in a cylindrical or box geometry) 
involving deformation of a bulk sample in one direction,  while the lateral
boundaries of the system are fixed. This test is particularly suited for determining the 
poroelastic properties of granular materials  \cite{imole2013hydrostatic, imole2013force, imole2014micro, bandi2013fragility}.
While most uniaxial tests on dry bulk solids have 
been devoted to studying the relationship between pressure and density and the bulk long time consolidation behavior, the dynamics of the time-dependent phenomena has been 
less studied in experimental and practical
applications \cite{zetzener2002relaxation}. For example, in standard shear testers like the Jenike 
shear tester \cite{kamath1994flow} and the Schulze ring shear tester \cite{schulze2003time}, during yield stress measurements, the focus is usually not on the relaxation behavior. 
Considerable stress-relaxation of bulk materials can even disturb yield stress measurements.
Additionally, most cohesive contact models \cite{luding2008cohesive, tomas2000particle, tykhoniuk2007ultrafine,walton1995force} used in 
discrete element simulation of granular materials do not account for the time dependent relaxation behavior, 
similar to those observed in viscoelastic materials such as polymers \cite{gloeckle1991fractional,metzler2003fractional,friedrich1991relaxation},
gels \cite{winter1986analysis,chambon1986rheology}, in dielectric relaxation \cite{jonscher1999dielectric,jonscher1977universal}
and in the attenuation of seismic waves \cite{kjartansson1979constant}. For the improvement of both discrete element contact models
and constitutive macro models relating to cohesive powders, it is necessary to have an experimental and theoretical understanding of the stress response of cohesive materials under different loading conditions. 

For viscoelastic materials, the relaxation has been 
reported to imply a memory effect and can be described using convolution integrals 
transformed to their fractional form and containing a relaxation modulus that describes the response of the system to stress \cite{schiessel1995generalized}. 
For these materials, phenomenological models involving the combination 
of springs and dashpots, such as the Maxwell, Zener, anti-Zener, Kelvin-Voigt, and the Poynting-Thomson models have been developed 
(see Refs.\ \cite{kruijer2006modelling, bagley1986fractional, atanackovic2002modified, mainardi2010fractional} and references therein).
Even though stress relaxation has also been observed in granular media \cite{zetzener2002relaxation, schulze2003time, bandi2013fragility}, not much work has been done in providing 
a theoretical description of this phenomenon for granular materials.

In the present study, using two simple testers, we perform oedometric compression tests with the main goal of investigating the relaxation behavior of 
industrial powders at different stress levels under constant strain (volume). Another goal is to provide a quantitative comparison between the relaxation behavior as observed in two
testers, namely the lambdameter \cite{kwade1994determination_1, kwade1994determination_2, kwade1994design} and the FT4 Powder Rheometer \cite{freeman2007measuring}, 
in order to confirm that this behavior occurs irrespective of the device used. The lambdameter has the peculiar advantage that 
both vertical and horizontal stress can be obtained simultaneously -- unlike the FT4 Powder Rheometer and other simpler oedometric test setups.  Finally, we will propose a simple model for 
stress relaxation that captures the relaxation of cohesive powders at different compaction levels.

The work is structured as follows: In section \ref{sec:characters}, we provide a characterization of the material sample, and in section \ref{sec:expsetup} the description of the experimental devices and the test protocols. 
In section \ref{sec:creeptheory}, we present the theoretical model for stress relaxation. Section \ref{sec:resultsdiscs} is devoted to the discussion of experimental and theoretical results, while the conclusions 
and outlook are presented in section \ref{sec:creepconclusn}.

\section{Sample Description and Material Characterization}   
\label{sec:characters}

\begin{figure}[!ht]
 \centering
\subfigure[]{\includegraphics[scale=1.3,angle=0]{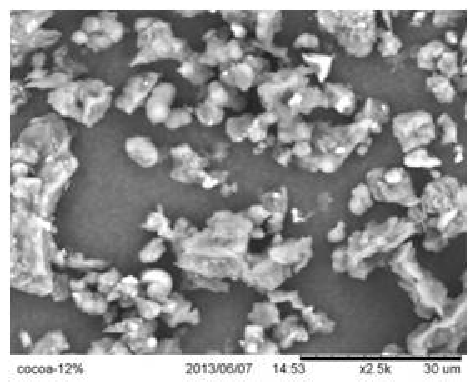}\label{cocoa12}}
\subfigure[]{\includegraphics[scale=1.3,angle=0]{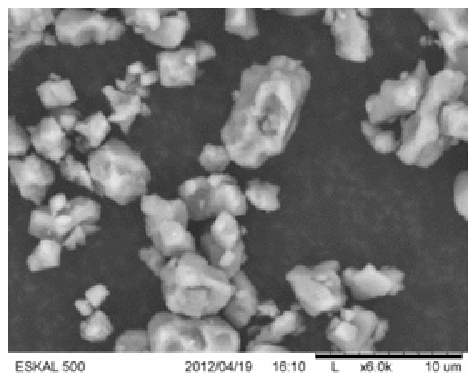}\label{eskal500}}
\caption{Scanning electron microscope images of the cohesive samples (a) Cocoa  with 12\% fat content  (b) Eskal 500 limestone powder. Note the different scales at the bottom right.}
\label{morphology}
\end{figure}

In this section, we provide a brief description of the experimental samples along with their material properties. 
In order to investigate the relaxation behavior, two cohesive reference samples were chosen, namely cocoa powder and Eskal 500 limestone. The choice is based on several selection
factors, among which are the suitability for different industrial applications, ability to withstand repeated loading without changes in the property of the sample and long term
availability/storage of the samples. The Eskal limestone has been used extensively as reference cohesive powder, and is made available in convenient amounts in a collaborative European project,
c.f.\ www.pardem.eu \cite{thakur2013characterization}. Scanning Electron Microscope (SEM) images obtained using a Hitachi TM 1000 Instrument (Hitachi Ltd, Japan) for both powders are 
displayed in Fig.\ \ref{morphology}. 

The particle size distributions are
measured using the Helos testing instrument (Sympatec GmbH, Germany). 
While limestone powder  is dispersed with air pressure, we use the wet mode to disperse cocoa powder since it forms agglomerates.
For the wet mode, cocoa powder is dispersed in dodecane, an oily liquid, in order to retain the fat layer while ultrasound (vibration) is applied to stress the dispersion and break off the agglomerates.
The particle density is measured by helium pycnometry (Accupyc, Micromeritics, USA) while the water content is given as the ratio of the difference between the original and dried mass 
(after 24 hours in a oven at 100$^\circ$C) and the original sample mass. 
The bulk cohesion is the limiting value of shear stress for which the normal stress is equal to zero and is determined from shear experiments with a Schulze ring shear 
tester (RST-01.pc by Dietmar Schulze Sch{\"u}ttgutmesstechnik, Germany).
A more specific description of the experimental samples is provided in the following section.

\subsection{Cocoa Powder}

One cohesive sample used in this work is cocoa powder with 12\% fat content - which is a representative sample for the material used as basic ingredient 
in the production of chocolate and related beverages. 
The material properties including size distribution, particle density and water content are shown in Table {\ref{tablematerialppt}}
along with a scanning electron microscope visualization of its morphology in Fig.\ \ref{cocoa12}. We note that even though the powder is relatively hygroscopic, 
its humidity does not change significantly during the experiments. Additionally, the experiments are performed over a relatively short period under ambient 
conditions and samples are sealed in air-tight bags when not in use to minimize effects that could arise due to changes in the product humidity.

\begin{table}
\caption{Material parameters of the experimental samples. }
\centering
 \begin{tabular}{lrlll}
    \hline
     \bf{Property} &  & \bf{Unit} & \bf{Cocoa (12\%)} & \bf{Eskal 500 limestone} \\ [0.4ex]
    \hline 
    Size distribution & $D_{10}$& $\mu$m & 2.14 & 1.34  \\ [0.4ex]
     &  $D_{50}$ & $\mu$m & 9.01 &  4.37  \\ [0.4ex]
    & $D_{90}$ & $\mu$m  & 37.40  &  8.24  \\ [0.4ex]
     Particle density &  &[kg/m\textsuperscript{3}] & 1509 & 2710  \\ [0.4ex]
    Water content  &  & $\%$ & $<$ 1.5\% &  $<$ 0.2\% \\ [0.4ex]
   Bulk cohesion (as function & $\sigma_c$ & ${\mathrm{kPa}}$ & 1.8 at 7.4 kPa & 1.3 at 4.6 kPa \\ [0.4ex]
    of major principal stress)&  &  & 9.6 at 41.8 kPa &  3.3 at 12.7 kPa  \\ [1ex]
    \hline
  \end{tabular}
  \label{tablematerialppt}
\end{table}

\subsection{Eskal 500 Limestone}

The other industrial powder sample used in this work is Eskal 500 limestone powder (KSL Staubtechnik, Germany). Eskal 500 limestone is a commercially available powder that has wide 
applications in architecture, road construction, blast furnaces, medicines and cosmetics. It  is also considered a suitable reference material 
for calibration and standard testing \cite{feise1998review, kwade1994design, zetzener2002relaxation}. One advantage of this material over other grades is its 
inability to absorb humidity from air. During long term storage under stress, Eskal 500 limestone shows no degradation as confirmed by repeatable results from experiments carried out 
under different conditions. The material properties and SEM morphology are shown in Table \ref{tablematerialppt} and Fig.\ \ref{eskal500}.

Comparing the physical features of the powders, cocoa powder is brownish while Eskal is whitish in color. Secondly, 
while cocoa powder contains some 12\% fat, Eskal 500 limestone does not. This distinction is
 important for a comparison of their relaxation behavior.

\section{Experimental Setup}
\label{sec:expsetup}
In this section, we describe the lambdameter and FT4 Powder Rheometer along with the protocols used in performing the tests.  

\subsection{FT4 Powder Rheometer}
\label{sec:ft4desc}

\begin{figure}[!ht]
 \centering
\subfigure[]{\includegraphics[scale=0.70,angle=0]{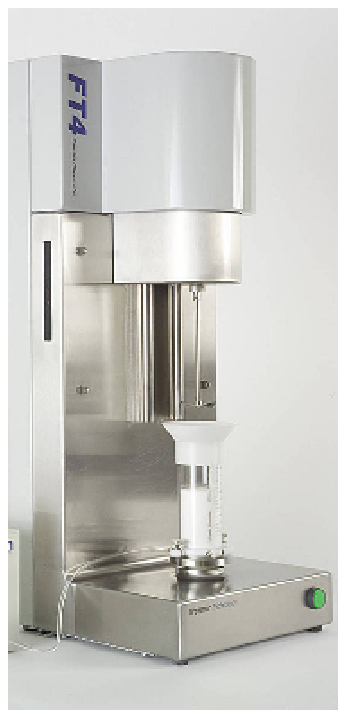}\label{ft4}}
\subfigure[]{\includegraphics[scale=0.335,angle=0]{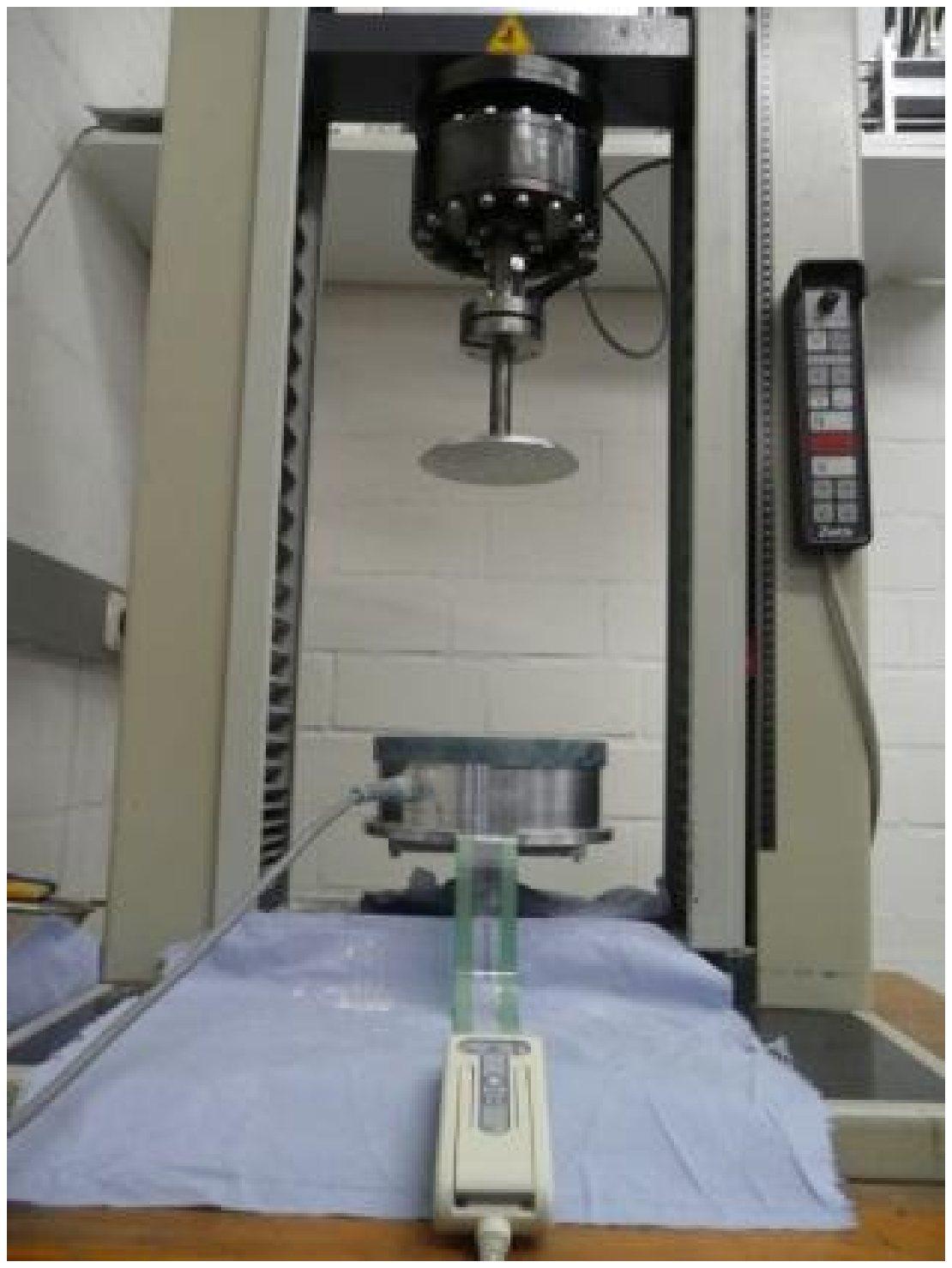}\label{lambda}}
\subfigure[]{\includegraphics[scale=0.33,angle=0]{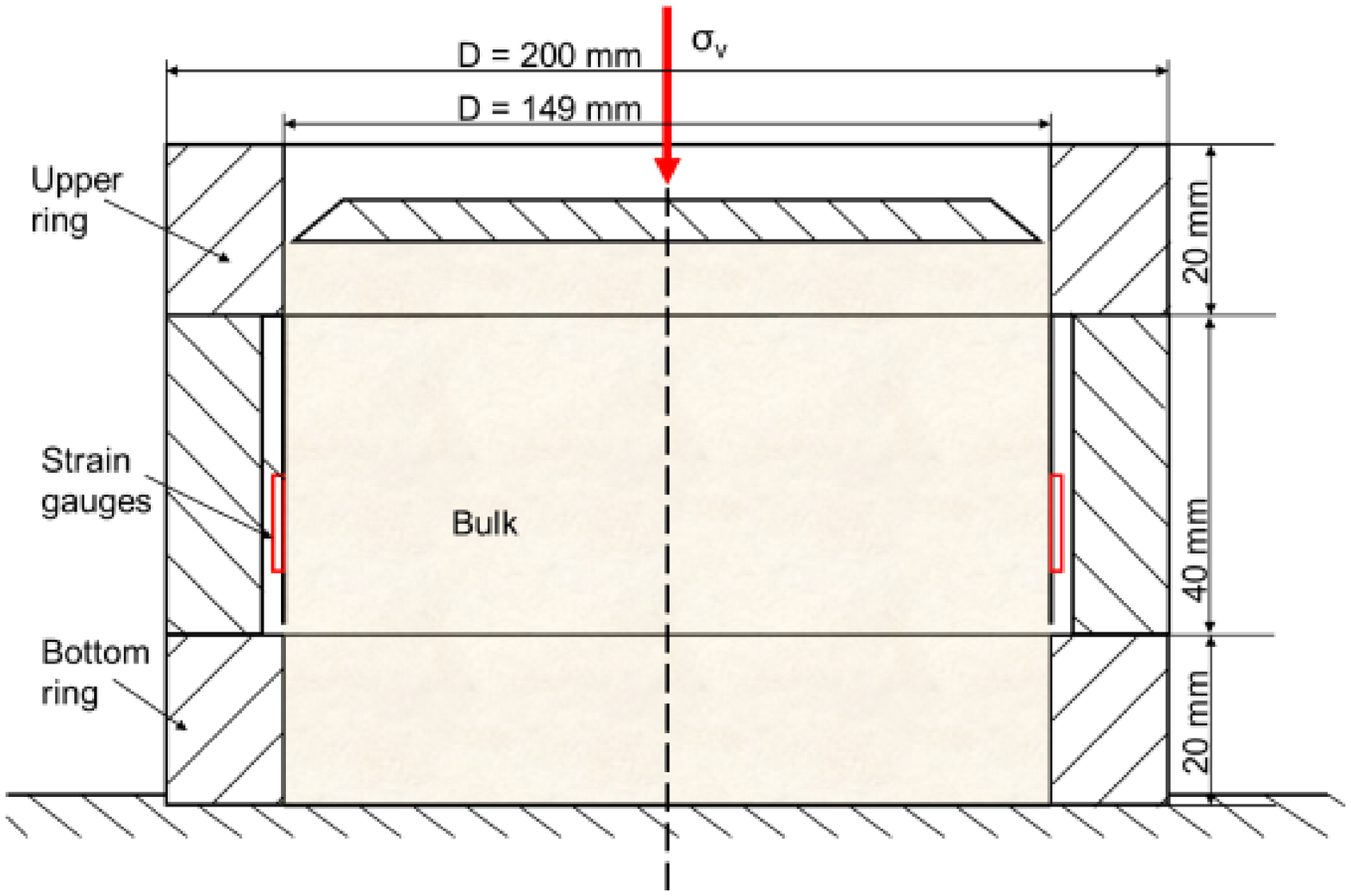}\label{schemalambda}}\\
\caption{(a) The FT4 Powder Rheometer \cite{freeman2007measuring} and (b) The lambdameter apparatus used for the experimental tests. (c) A schematic representation of the lambdameter test set-up.}
\label{schematics}
\end{figure}

The first experimental equipment used in this work is the FT4 Powder Rheometer (Freeman technology Ltd. UK), illustrated in Fig.\ \ref{ft4}.
Standard accessories for the compressibility test include the 50 mm diameter blade for conditioning, the vented piston for compression and the 50 mm height by 50 mm diameter borosilicate test vessel. 
One advantage of the commercial FT4 Powder Rheometer is the automated nature of the test procedure requiring minimal operator intervention. 

The compression test sequence is as follows: 
The sample is placed in the test vessel after the tare weight of the vessel has been obtained. 
The weight of the powder is measured and the conditioning cycle is initiated. 
The conditioning procedure involves the gentle movement of the conditioning blade into the test sample to gently disturb the 
powder bed for a user pre-defined number of cycles. This action creates a uniform, 
lightly packed test sample that can be readily reproduced. In this study, we allow three pre-conditioning cycles before the uniaxial 
compression tests are carried out. 
Subsequently, the blade is replaced with a vented piston, which incorporates a stainless steel mesh to allow the entrained air in the powder to escape uniformly across the surface of the powder bed. 
The vessel assembly is split (or leveled) to provide precise volume measurement and the powder mass is recalculated after splitting. The compression test then begins with the distance 
travelled by the piston measured for each applied normal stress.

\begin{table}
\caption{Comparison of the FT4 Powder Rheometer and the lambdameter specifications.}
\centering
 \begin{tabular}{lrrrrr}
    \hline
    \bf {Property} & \bf{FT4 Rheometer} & \bf{Lambdameter}  \\ [0.4ex]
    \hline \\
    Cell volume & 8.5 $\times 10^{-5}$ m$^3$ & 1.39 $\times 10^{-3}$ m$^3$\\ [0.4ex]
    Cell shape & cylindrical & cylindrical \\ [0.4ex]
    Wall material & borosilicate glass & aluminium alloy \\ [0.4ex]
    Diameter (D) & 0.05 m & 0.149 m \\ [0.4ex]
    Height (H) & 0.02 m, 0.05 m & 0.08 m  \\ [0.4ex]
    Aspect ratio $\alpha = H/D$ & $0.4, 1$ & $0.53$  \\ [0.4ex]
    Driving mode & motor control & motor control \\ [0.4ex]
    Test  control & built in test program on PC & Labview \\ [0.4ex]   
   Sample weighing & on-board & offline\\ [0.4ex]
    Compression device & vented piston & top plate \\ [0.4ex]
    Driving velocity & variable & variable  \\ [0.4ex]   
    Maximum stress & 22 kPa &69.96 kPa\\ [0.4ex]
    Sample pre-conditioning & automatic & manual\\ [0.4ex]
    Test duration & variable & variable\\ [0.4ex]
    Stress measurement & vertical stress  & horizontal and  \\ 
    (direction) & & vertical stress \\ [0.4ex]
  \hline
  \end{tabular}

  \label{comparelambdaft4}
\end{table}

\subsection{The Lambdameter}

The custom made lambdameter represents a horizontal slice of a silo and is primarily used in obtaining the lateral stress ratio \cite{kwade1994determination_1, kwade1994determination_2, kwade1994design} -- one 
of the most important parameters in the calculation of stress distributions
in silos for reliable design \cite{schwedes1968fliessverhalten, schwedes2003review}.
The lambdameter used was designed at the Institute for Particle  Technology (iPAT), Technische Universit{\" a}t Braunschweig, and is shown schematically in
Fig.\ \ref{lambda} and \ref{schemalambda}. The lambdameter measures the vertical (axial) and horizontal (radial) stress of a powder under compaction. The horizontal stress is measured through the installation of pressure
cells along the periphery of the cylindrical mould. The measuring ring of the lambdameter is made from aluminium alloy with a very smooth surface and the dimensions are listed in  Table \ref{comparelambdaft4}. 
To allow for the automation of the compression test, similar to that of the FT4, the lambdameter setup is installed into a 
Zwick Z010 (Zwick/Roell, Zwick GmbH \& Co. KG, Germany) uniaxial testing device as shown in Fig.\ \ref{lambda}.

The experimental procedure is as follows. The experimental sample is first sieved to prevent formation of agglomerates and a spoon is used to fill the sample evenly into the 
cylindrical mould until it is completely full. Using a smooth object, 
excess material is removed without allowing for a compaction of the sample in the mould. 
Next, the top plate lowers according to a prescribed velocity and the force-displacement measurement is initiated once a contact force is detected.
The horizontal and vertical stresses along with the position of the top punch at a given 
time are recorded with a data logger on a computer connected to the experimental setup.

A detailed comparison between the main features of both testers is shown in Table \ref{comparelambdaft4}. We only note that the FT4 is more automated and requires less human intervention compared to the lambdameter.
On the other hand, in contrast to the FT4 Powder Rheometer, higher volumetric strains and axial stresses can be reached with the lambdameter. Additionally, the lambdameter also provides for the measurement of the 
horizontal (lateral) stress.

\subsection{Test Protocols}

In order to investigate the relaxation behavior of the different experimental samples under uniaxial loading, different staged test protocols are employed, see Table \ref{protocoltable}.
The uniaxial loading is done in steps of 5 kPa with intermediate relaxation between each step. The maximum stress reached for experiments with
the FT4 Powder Rheometer is 22 kPa while for the lambdameter, a higher maximum stress of 25 kPa is reached. 
We performed more 
extensive experiments with the lambdameter setup due to its versatility in terms of the maximum stress reached and the horizontal stress measurement. We 
compare in some cases the stress-relaxation behavior under uniaxial loading for both equipments. 
Using protocol 1 with Eskal 500 limestone as sample, three tests were performed to investigate the reproducibility of the FT4 and two tests for the lambdameter measurements. 
The results were found to be reproducible up to 3--5 percent in the FT4 while the reproducibility
was 6 percent in the lambdameter. 

\begin{table}
\small
\caption{Table of experimental protocols performed. Note that for experiments with the 
FT4 Powder Rheometer, the maximum stress reached is 22 kPa. Protocols 1--4 represent
a variation of the  relaxation time, while 5--9 are different compression rates. Crosses (x) indicate the device used in performing the 
experiment.}
\centering
 \begin{tabular}{lrrrcrl}
    \hline
  \bf{ Protocols} & \bf{Velocity} & \bf{No. of} & \bf{Vertical}  & \bf{Relaxation} & \bf{FT4} &  \bf{Lambda-} \\
      &  \bf{[mm/s]} & \bf{steps} & \bf{stress [kPa]} &  \bf{time [mins]}  &  & \bf{meter}\\ [0.4ex]
    \hline 
   
   Protocol 1 & 0.05 & 5 & 5 - 10 - 15 - 20 - 22/25 & 5 & x & x\\ [0.4ex]
    Protocol 2 & 0.05 & 5 & 5 - 10 - 15 - 20 - 25  & 10  &  & x\\ [0.4ex]
    Protocol 3 & 0.05 & 5 & 5 - 10 - 15 - 20 - 25 & 20 &  & x \\ [0.4ex]
     Protocol 4 & 0.05 & 5 & 5 - 10 - 15 - 20 - 25  & 30 & &  x\\ [0.4ex]
    \hline
     Protocol 5 & 0.01  & 5 & 5 - 10 - 15 - 20 - 22/25 & 10 & x & x\\ [0.4ex]
    Protocol 6 & 0.3  & 5 & 5 - 10 - 15 - 20 - 25 & 10 &  & x \\ [0.4ex]
    Protocol 7 & 0.7 & 5 & 5 - 10 - 15 - 20 - 25 & 10 &  & x \\ [0.4ex]
    Protocol 8 & 1.0 & 5 & 5 - 10 - 15 - 20 - 25 & 10 &  & x \\ [0.4ex]
    Protocol 9 & 1.3 & 5 & 5 - 10 - 15 - 20 - 25 & 10 &  & x \\ [0.4ex]
  \hline
  \end{tabular}
  \label{protocoltable}
\end{table}

In general, we study the effects of strain rate, relaxation time duration and the stress at which the relaxation is initiated along the loading path. We measure the vertical stress as function of the volumetric strain 
i.e. vertical strain since we are in an oedometric setup.

\section{Stress Relaxation Theory}
\label{sec:creeptheory}

Assuming a vertical stress $\vstress = f/A$, which acts on the top plate under uniaxial loading, the change of stress with time, i.e. the 
stress-rate should be higher for stronger applied stress due to a micro- or nano-scopic change of the contact structure. 
The model evolution relation is:

\begin{equation}
 \displaystyle \frac{\partial}{\partial t} \vstress = -\frac{C}{t_0+t}\vstress,
\end{equation}
where $C$ is a dimensionless proportionality constant and $t_0$ is a typical response time. The time $t$ in the denominator on the right hand side accounts for the fact that the change of stress
decays with time extremely slowly. 
In order to visualize the model, consider that for organic materials like coffee and cocoa, the initial grains contain liquid and solid ingredients. 
Due to strong stresses, the liquid is squeezed out of the solid matrix -- locally at contacts that experience strong forces. The terminal 
state would be a state where all liquid content has been squeezed out, however, since pores exist on many scales, this can take extremely long, i.e.\ much longer than the experiments which were performed here.
The stress-rate is also proportional to the stress itself, since at zero stress, there is no reason to assume further stress changes. 

The constant $C$ determines the magnitude of the stress-rate and contains information about the
microscopic constitution and composition of the material. Hard materials with low liquid content are described by large $C$ values, whereas soft materials with high liquid content correspond to small $C$ values.

Assuming that the stress is raised from zero to a value $\vstressmax$ instantaneously the response of the system is then given by the solution of the above equation with initial stress $\vstressmax$ and starting from time $t$=0, so that:

\begin{equation}
\displaystyle \ln \left( \frac{\vstress}{\vstressmax} \right) = \int^{\vstress}_{\vstressmax}  \frac{\partial \vstress'}{\vstress'} =  \int^t_{0} - \frac{C}{t_0 + t'} {\partial t'} =
-C \ln(t_0 + t) + C \ln t_0 = \ln \left ( \frac{t_0 + t}{t_0} \right )^{-C}
\end{equation}
which can be further simplified to:

\begin{equation}
 \frac{\vstress}{\vstressmax} = \left ( \frac{t_0 + t}{t_0} \right )^{-C} =  \left ( 1 + \frac{t}{t_0} \right )^{-C}
\label{eq:relaxmodel}
\end{equation}

Note that the model can also be formulated in terms of the force rate \cite{verzijden2005flow}. In the next section, the simple model presented above will be compared to experimental 
data and the response time $t_0$ and the parameter $C$ will be analyzed.

\section{Results and Discussion}
\label{sec:resultsdiscs}

In this section, as results of the current study, we will first compare the uniaxial compression and relaxation experiments
carried out with the FT4 Powder Rheometer for different aspect ratios. To complement these results, we also discuss the dependence on sample material characteristics. 
Finally, we investigate the effects of strain-rate, loading steps and relaxation time on the decay of the stress
at constant strain.

\subsection{Dependence on Aspect Ratio}
\label{sec:compare_equip_aspectratio}

In order to investigate the role that different vessel aspect ratios play in the stress-strain evolution, using the FT4, we perform uniaxial compression tests on cocoa powder
with a carriage speed of 0.05 mm/s (protocol 1 in Table \ref{protocoltable}). Two aspect ratios ($\alpha$) are considered namely, $\alpha =$ 0.4 and 1.0. 
For these tests, the vessel diameter $D =$ 0.05 m is fixed  while the filling height $H$ of the vessel is changed from 0.05 m to 0.02 m to achieve the target aspect ratios. 
Additionally, five intermediate relaxation stages (R1--R5),
in which the top piston/punch is held in position 
for 300 seconds at specific intervals of 5 kPa are included during the compression test.

\begin{figure}[!ht]
 \centering
\subfigure[]{\includegraphics[scale=0.37,angle=270]{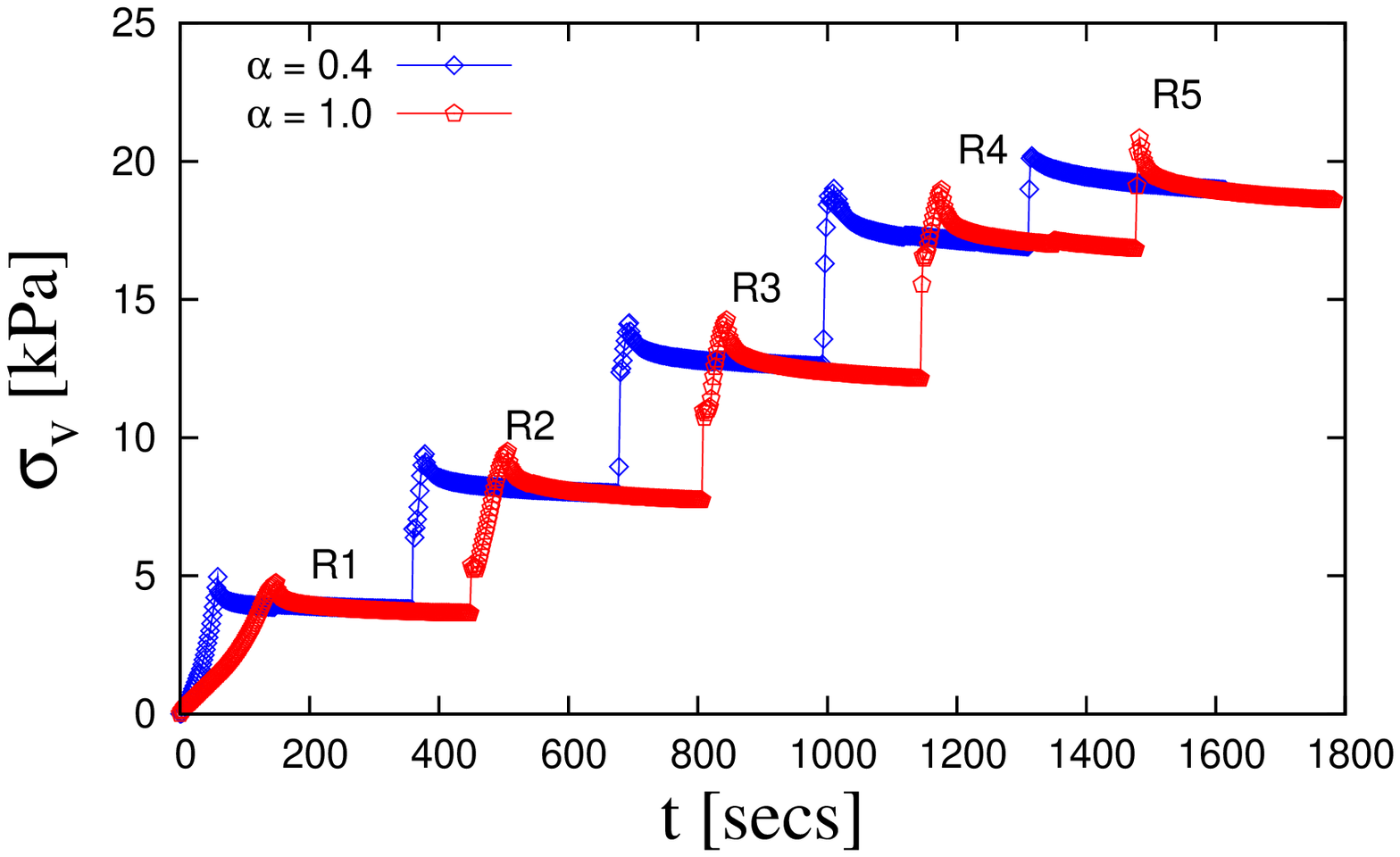}\label{comp_ft4_asp_time}}
\subfigure[]{\includegraphics[scale=0.37,angle=270]{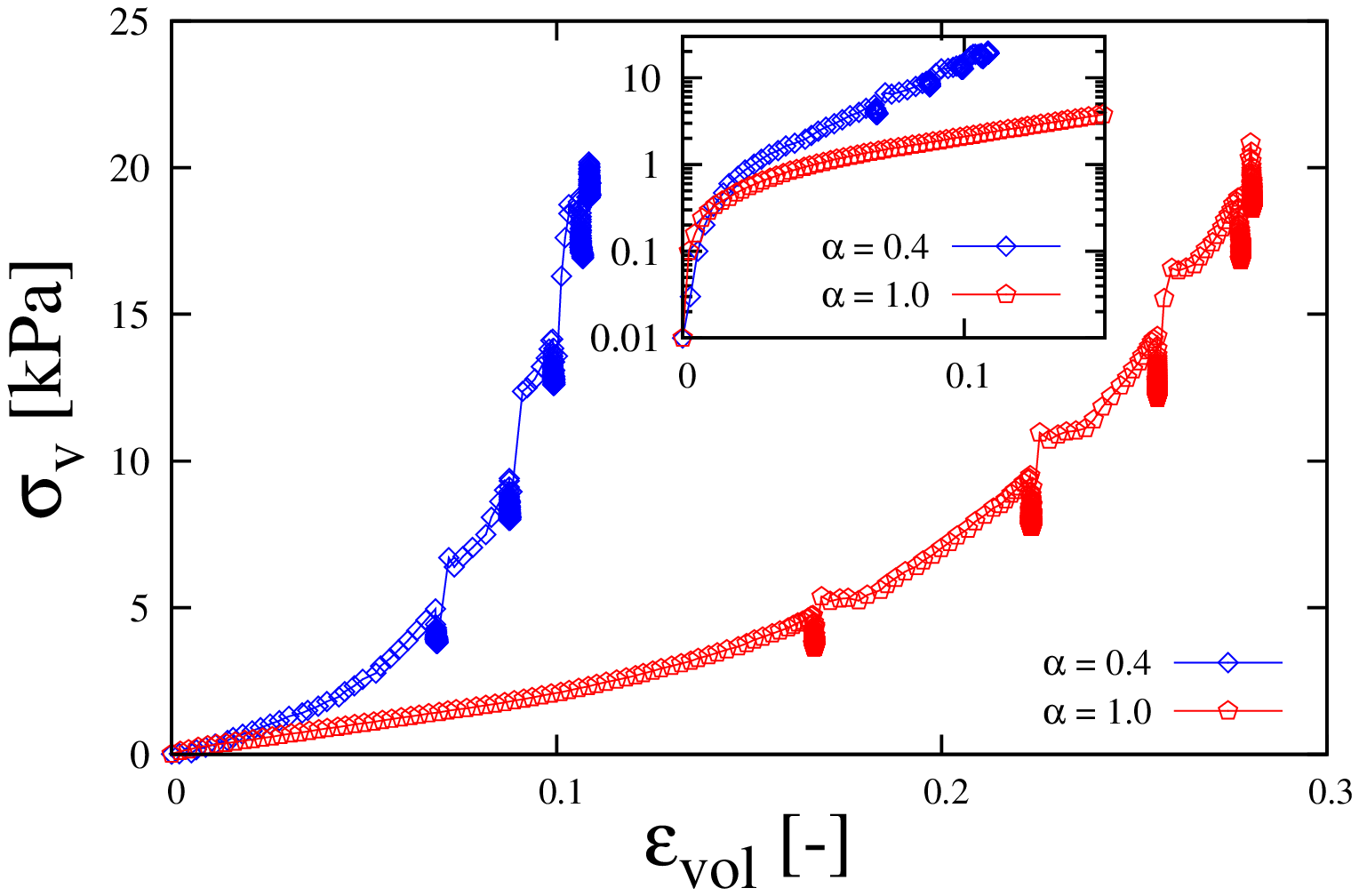}\label{comp_ft4_strain}}
\caption{Comparison of the vertical (axial) stress plotted against (a) time (b) volumetric strain for experiments with cocoa powder and Eskal 500 limestone. Experiments carried out
using the lambdameter with carriage velocity 0.05 mm/s, while R1--R5 represent the intermediate relaxations during loading.}
\label{compare_asp}
\end{figure}

In Fig.\ \ref{comp_ft4_asp_time}, we plot
the vertical stress $\vstress$ as function of time. During loading, the axial stress builds up with time until the first target stress of 5 kPa (at R1) is reached.
We observe a slower increase of the axial stress with time for the higher aspect ratio even though the respective pistons were moved with the same 
speed. Consequently, the vertical compaction and thus strain at constant vertical stress is higher in the setup with $\alpha =$ 1.0 than in the setup with $\alpha =$ 0.4. 
This is possibly due to the difference in  sample masses for both equipments, 
leading to different initial densities of the same sample, and consequently producing different response to compressive stress. Other possible reasons will be discussed below.

With the initiation of the first relaxation R1 at 5 kPa, we observe for both aspect ratios a time-dependent stress relaxation during the rest-time of 300 seconds. 
This observation has been reported in literature for other test setups and granular materials \cite{schulze2003time, zetzener2002relaxation}, confirming that the stress relaxation is not due to a drift in the measuring equipments but it is 
a material feature happening at their contacts. Other reasons for the relaxation of the powder under stress are the escape of air trapped inside 
the bulk pores during (fast) compression and a successive relocation of particles resulting in the ``softening'' of the vertical
stresses.

The activation
of axial compression after relaxation leads to a sharp increase in the axial stress until the next intermediate stress state is reached. This sudden jump is similar to that observed
in stick/slip \cite{schulze2003time, shinbrot2012electrostatic} experiments  and friction between solid bodies \cite{dieterich1994direct} 
where a sudden increase in shear velocity results in a sharp increase in shear stress. The same features are reproduced for higher stress states.

An objective comparison of the stress-strain evolution in both testing equipments is presented in  Fig.\ \ref{comp_ft4_strain}, where the vertical stress is plotted against volumetric strain. The volumetric strain is defined
here as $ \evol = -(L-L_0)/L_0$ where $L$ and $L_0$ are the actual and initial piston positions, respectively.

At the initial stage, the response to applied stress for both aspect ratios is almost identical. Shortly afterwards, the dependence on aspect ratio
kicks in and the setup with the higher aspect ratio ($\alpha =$ 1.0)  produces a softer response to the applied stress in comparison to $\alpha =$ 0.4.   
During the relaxation phase, the decrease in stress occurs at constant strain as shown by the vertical drops along the deformation path for both aspect  ratios.

The softer response observed for the higher aspect ratio 
can also be explained by the difference in sample height which, according to 
Janssen \cite{janssen1895versuche}, causes weaker stress far away from the piston for $\alpha =$ 1.0. 
It follows that the stress-strain evolution in the two different setups is influenced by the difference in aspect ratio, where increasing the sample height leads to slower and softer stress-strain response.

\subsubsection*{Relaxation Steps}
\label{sec:compare_relax_aspect}

\begin{figure}[!ht]
 \centering
\subfigure[]{\includegraphics[scale=0.37,angle=270]{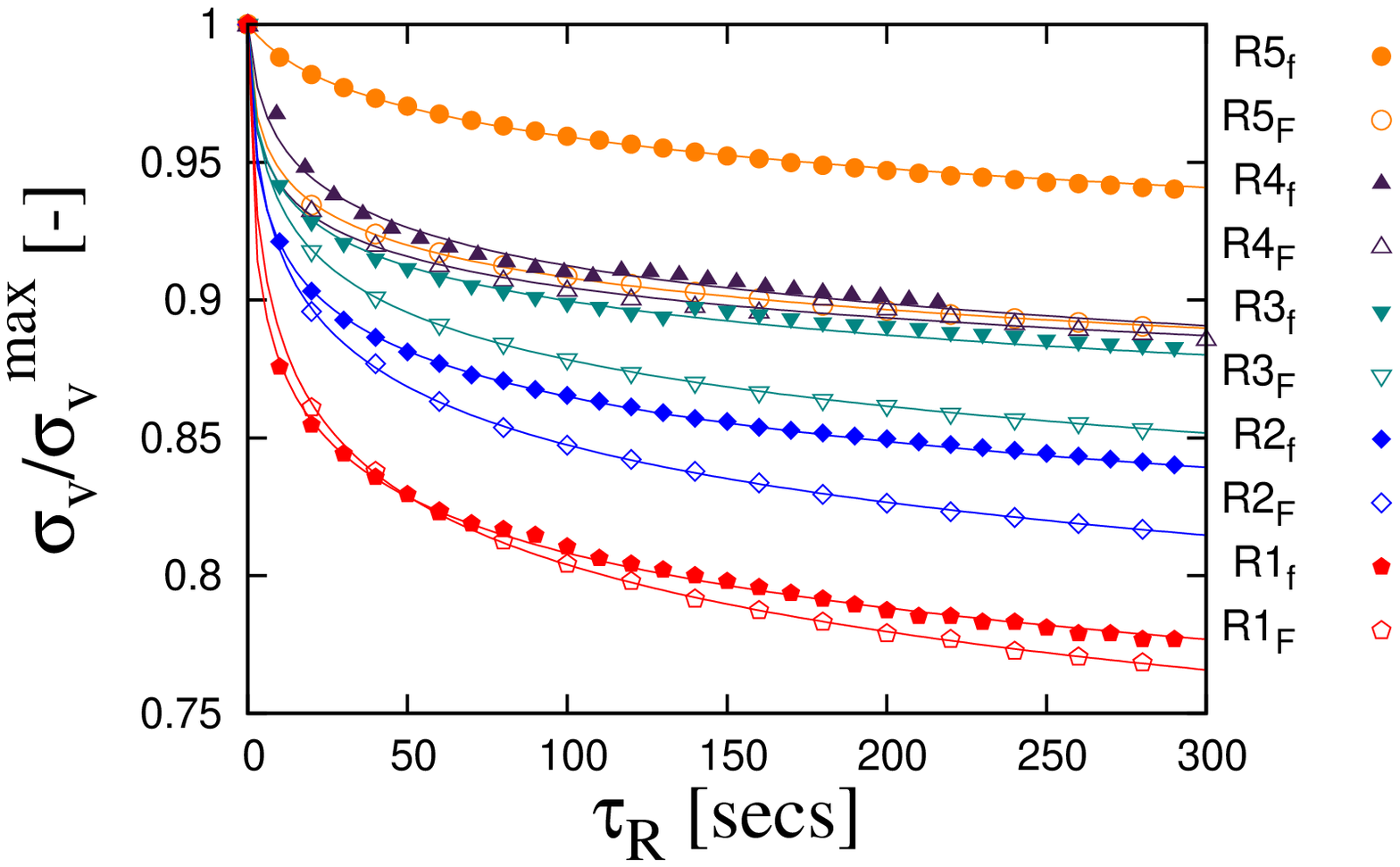}\label{asp}}
\subfigure[]{\includegraphics[scale=0.32,angle=270]{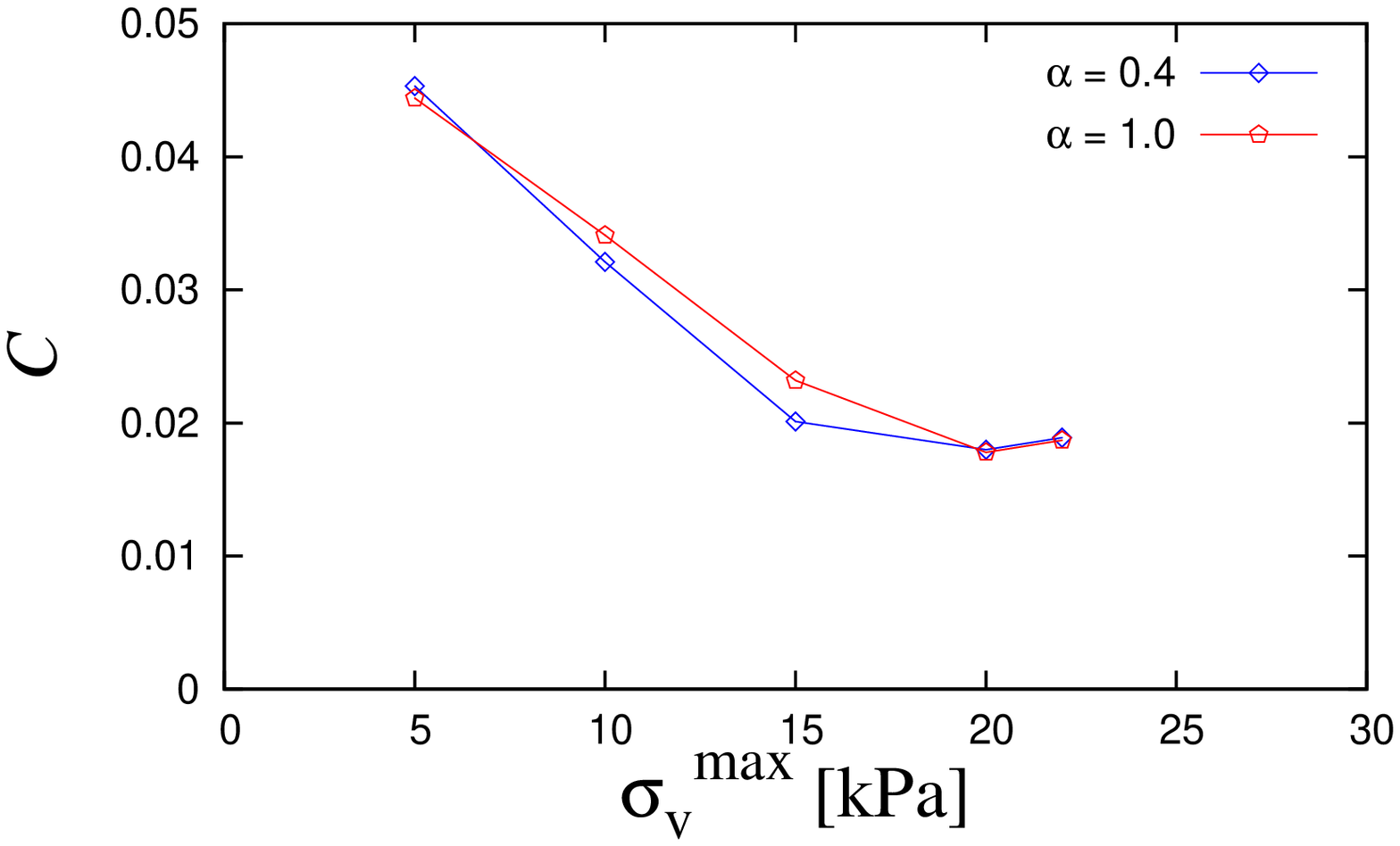}\label{asp_cvals}}
\caption{(a) Relaxation at different stress levels R1--R5 during the uniaxial compression of cocoa powder in Fig.\ \ref{compare_asp}. The subscripts f (solid symbols) and F (open symbols)
represent data for $\alpha =$ 0.4 and 1.0, respectively.  The symbols represent experimental data while the solid lines represent the theoretical fit using Eq.\ (\ref{eq:relaxmodel}) with parameters listed in Table\ \ref{parametertablea}. 
Even though the data output was at 50 Hz, we show only points at intervals of $\approx$ 20 Hz to allow for a clear visualization of the relaxation process. (b) Dimensionless parameter $C$, as displayed in 
Table\ \ref{parametertablea}, plotted as function of stress level.}
\label{ratio}
\end{figure}

Next, we turn our attention to the relaxation stages and extract from Fig.\ \ref{compare_asp} the data for the steps R1--R5 plotted in Fig.\ \ref{asp}. For clarity, R1 is termed the first relaxation occurring at $\approx$ 5 kPa while R5 is the final 
relaxation at 21 kPa.  The vertical stresses have been normalized by their initial values before relaxation  while $\tr$ is the relaxation time which in this case is limited to 300 secs.

In general, we consistently observe stronger relaxation amplitudes for earlier, lower stress relaxations and considerably stronger relaxation in the test with the higher aspect ratio (open symbol). 
The stress relaxation law proposed in Eq.\ (\ref{eq:relaxmodel}) describes well the relaxation for all aspect ratios at all stress states 
after the stress in each state has been normalized by its maximum value $\vstressmax$ such that $\vstress^* = \vstress/\svmax $ has a maximum value of 1. The maximum intermediate stress reached
after 300 s and the other parameters are displayed in Table\ \ref{parametertablea}. 

\begin{table}
\caption{Fit parameters for the analytical predictions of the relaxation model Eq.\ (\ref{eq:relaxmodel}). The subscripts $F$ and $f$ represent 
data for aspect ratio $\alpha =$ 1.0 and 0.4, respectively while R1--R5 are the relaxation steps. }
\centering
 \begin{tabular}{l|rrr|rrr}
    \hline
    & $\smax_{F}$& $t_{0F}$ & $C_F$ & $\smax_{f}$ & $t_{0f}$ &  $C_f$ \\ [0.4ex]
    \hline 
	 Step & & $\alpha = $ 1.0 & & & $\alpha = $ 0.4 &  \\ [0.4ex]
    \hline 
   R1  & 4.75 & 0.556 & 0.0444 & 4.75 &  0.2746 & 0.0453\\ [0.4ex]
    R2 & 9.5 &1.0308 & 0.0341 & 9.51 & 0.8209 & 0.0321\\ [0.4ex]
    R3 & 14.25 &1.0227  & 0.0212  & 14.26 & 0.5309 & 0.0201 \\ [0.4ex]
    R4 & 19.01 & 0.3614 & 0.0178 & 19.01 & 1.6542 & 0.0180\\ [0.4ex]
    R5 & 20.9 &0.5809 & 0.0187 & 20.93 & 12.4611 & 0.0189\\ [0.4ex]
    error [\%]  & -- & 0--5 & 0--1 & -- &  0--1 &0--0.5 \\ [0.4ex]
  \hline
  \end{tabular}
  \label{parametertablea}
\end{table}

Comparing the parameters of the model, we observe that the response time $t_0$ fluctuates, especially for the lower aspect ratio,  with no clear trend for increasing stress 
from R1--R5. On the other hand, the 
parameter $C$ displayed in Table\ \ref{parametertablea} and plotted in Fig.\ \ref{asp_cvals} shows a consistently decreasing trend  
and the values are close for both aspect ratios. This suggests
that $C$ is a material parameter that is not influenced by the aspect ratio of the experimental setup used, but by the stress and thus also by the 
history of the sample.

\subsection{Dependence on Material Characteristics}
\label{sec:powder_comparison}

As a second step, in order to compare the response of different cohesive powders, we introduce the second powder (Eskal 500) and 
repeat the same protocol as described in section\ \ref{sec:compare_equip}. 
For the sake of brevity, the comparison is done using only the lambdameter setup.

\begin{figure}[!ht]
 \centering
\subfigure[]{\includegraphics[scale=0.37,angle=270]{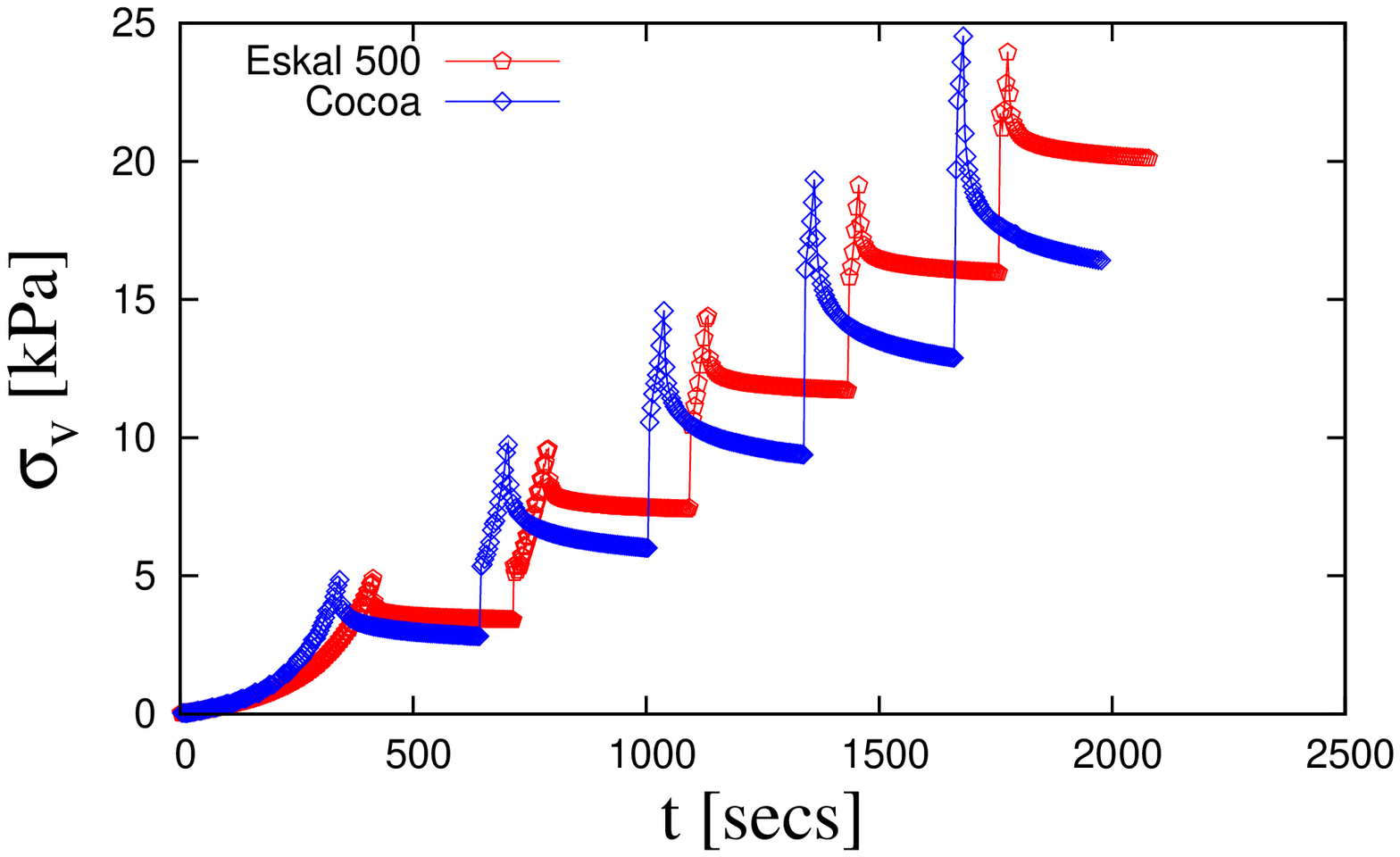}\label{comp_powd_time}}
\subfigure[]{\includegraphics[scale=0.37,angle=270]{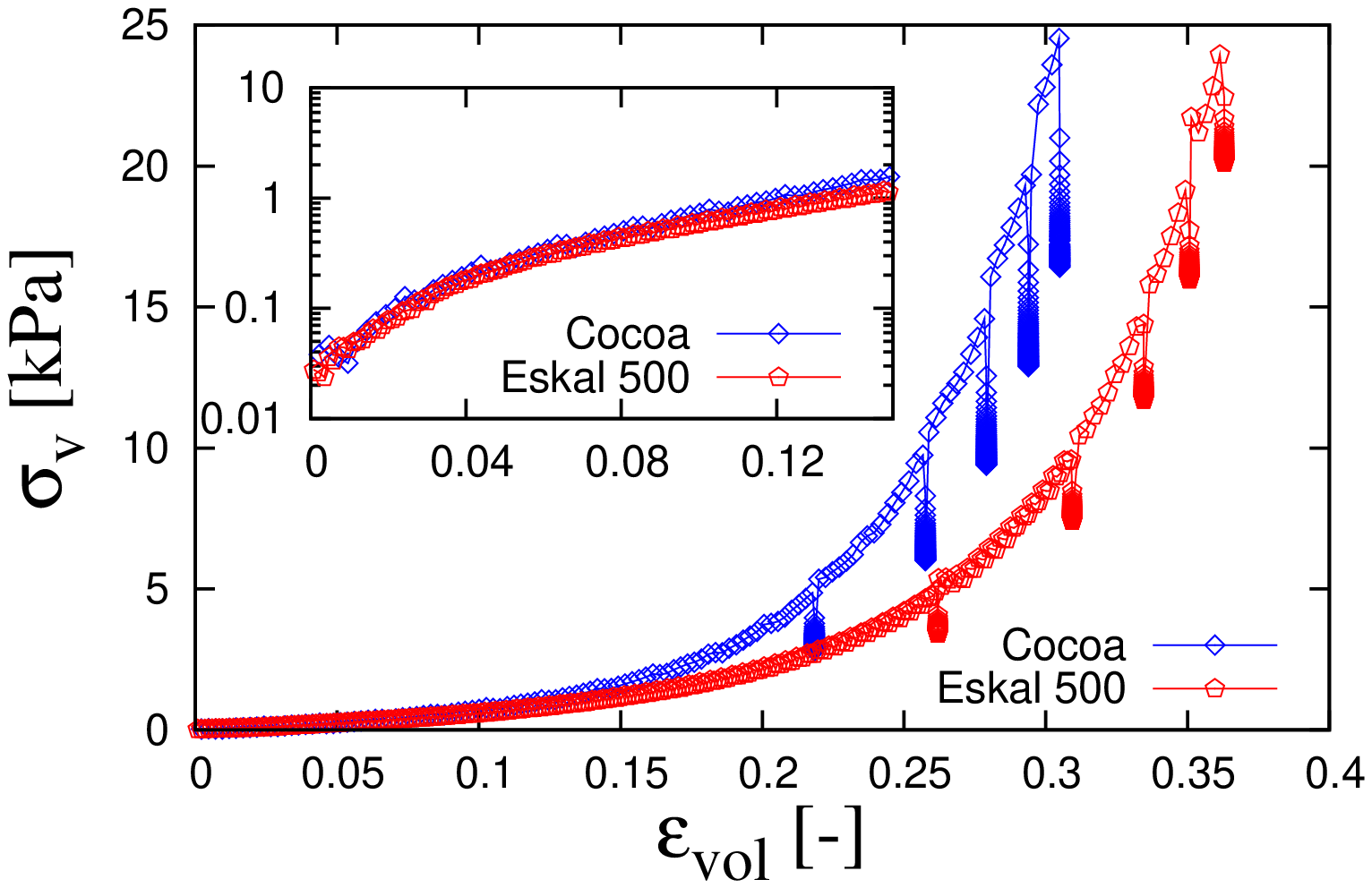}\label{comp_powd_strain}}
\caption{Comparison of the vertical (axial) stress plotted against (a) time (b) volumetric strain for experiments with cocoa powder and Eskal 500 limestone. Experiments carried out
using the lambdameter with carriage velocity is 0.05 mm/s while R1--R5 represent the intermediate relaxation steps during loading.}
\label{compare_powders}
\end{figure}

In Fig.\ \ref{comp_powd_time}, we show the time evolution of stress during compression for cocoa powder and Eskal 500 limestone.
Both powders show qualitatively identical relaxation behavior under applied stress. Comparing the stress-strain
response for both materials, as presented in  Fig.\ \ref{comp_powd_strain}, we observe
a similar response for both materials within the small strain region ($\evol < $  0.15). However, at larger strains 
the response diverges and limestone responds softer to strain, evidenced by the slower increase in vertical stress. Secondly, 
we confirm that the stress relaxation at R1--R5 occurs at 
constant strain as shown by the vertical drops along the deformation path. For the same intermediate stress, the onset of relaxation 
occurs at a higher strain in limestone as in cocoa.
We explain this differences by the finer particle size of limestone resulting in much more contact points, higher 
van der Waals forces between the particles, and higher air entrapment.

\begin{figure}[!ht]
 \centering
\subfigure[]{\includegraphics[scale=0.37,angle=270]{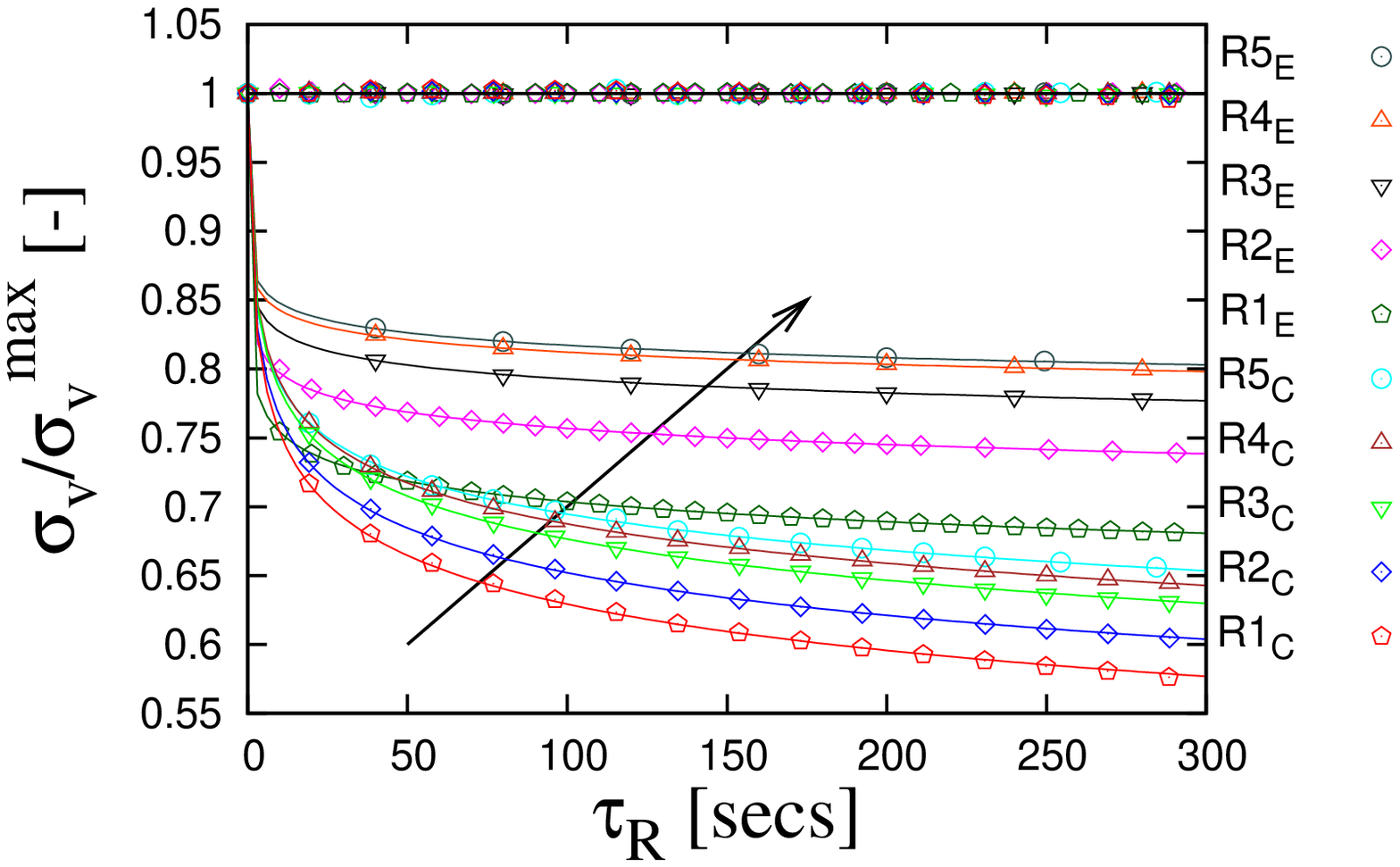}\label{comp_powd_relax}}
\subfigure[]{\includegraphics[scale=0.37,angle=270]{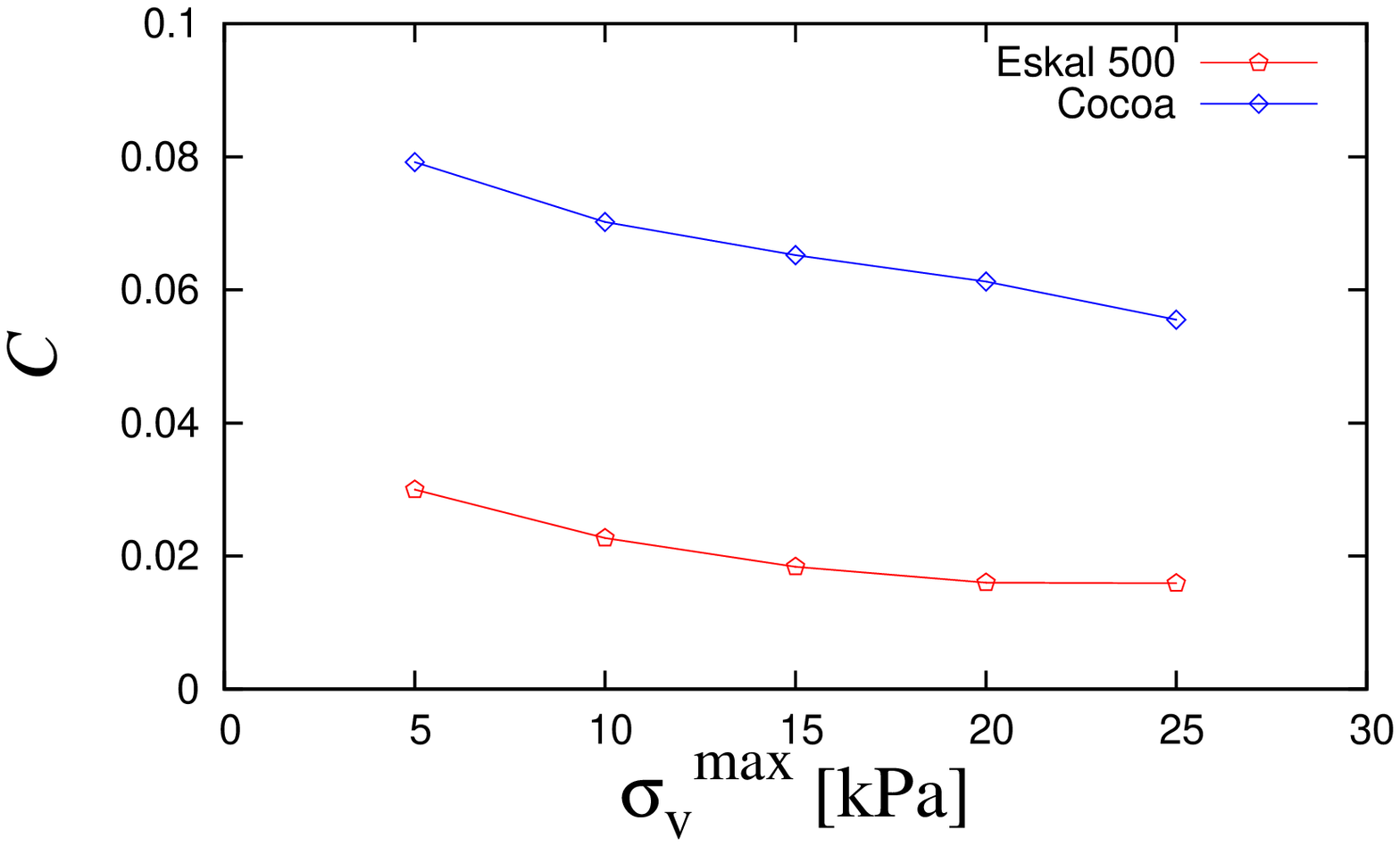}\label{powd_cval}}
\caption{(a) Extract of the 5 relaxation steps R1--R5  for both experimental specimens in Fig.\ \ref{compare_powders}. The subscripts E and $C$ 
represent data from experiments with Eskal 500 (E) and cocoa (C), respectively. The symbols represent the experimental data while
the solid lines represent the analytical Eq.\ (\ref{eq:relaxmodel}). The horizontal line is the quality factor -- experimental data divided by fit
function showing excellent agreement. (b) Evolution of the dimensionless parameter $C$ of Eq.\ (\ref{eq:relaxmodel}) with intermediate maximum stresses $\svmax$ for 
the two powders. }
\label{compare_powd_relax}
\end{figure}

In Fig.\ \ref{comp_powd_relax}, we extract the relaxation phases of the experiments shown in Fig.\ \ref{compare_powders} Eskal  (E) and cocoa (C) and plot them against the 
relaxation time. We observe that at the same stress and using the same driving velocity, cocoa powder relaxes more and much faster
than the Eskal limestone. For example, the relaxation under the lowest 
compressive stress (R1) shows a 33 percent decrease in stress for Eskal compared to a 43 percent decrease for cocoa. This is possibly arising from the fat content present in the cocoa powder.

\begin{table}
\caption{Fit parameters for the analytical predictions of the relaxation model Eq.\ (\ref{eq:relaxmodel}). The subscripts $C$ and $E$ represent 
data for cocoa and Eskal, respectively, while R1--R5 are the relaxation steps.}
\centering
 \begin{tabular}{lrrr|rrr}
    \hline
    Step & $\smax_{C}$ &$t_{0C}$ & $C_C$ & $\smax_{E}$& $t_{0E}$ & $C_E$ \\ [0.4ex]
    \hline 
   R1 & 4.8964 &0.28945 & 0.0792 & 5.0027 & 0.00082 & 0.0300 \\ [0.4ex]
    R2 & 9.9684 &0.2274 & 0.0702 & 10.0246 & 0.00046  & 0.0227 \\ [0.4ex]
    R3 & 14.9130 &0.2503 & 0.0652 & 15.0648 &0.00032  & 0.0184 \\ [0.4ex]
    R4 & 20.0375 &0.2207  & 0.0612 & 20.0004 & 0.00021 & 0.0160  \\ [0.4ex]
    R5 & 25.0718 &0.1422 & 0.0556 & 25.0253 &0.00032 & 0.0159\\ [0.4ex]
    error[\%] & --& 0--3 & 0--0.4 & -- & 3--7 & 0--1.3\\ [0.4ex]
  \hline
  \end{tabular}
  \label{parametertableb}
\end{table}

The fit parameters of Eq.\ (\ref{eq:relaxmodel}) are shown in Table \ref{parametertableb} for the five relaxation data 
depicted for each powder in Fig.\  \ref{comp_powd_relax}, limestone and cocoa.
The response time $t_0$ and dimensionless parameter $C$ for each powder generally show a decreasing trend with the maximum stress at which the relaxation is initiated.
The decreasing trend of both parameters $t_0$ and $C$ is confirmed also for Eskal 500, however, the time-scale is orders of magnitude smaller while
$C$ is of the same order only about a factor of two smaller, as summarized in table \ref{parametertableb} and plotted in Fig.\ \ref{powd_cval}.  

In summary, we conclude that even though both Eskal and cocoa powder show qualitatively similar relaxation at constant strain, their individual magnitudes and responses
are quantitatively dissimilar at different intermediate stress.

\subsection{Dependence on Relaxation Duration}
\label{sec:rtime}

In order to compare the changes in the vertical stress drop due to the relaxation duration, using the lambdameter, we perform several experiments in which the relaxation
time is varied between 5 minutes and 30 minutes (protocols 1--4 in Table\ \ref{protocoltable}). 
Afterwards, for each achieved stress state, only the first five minutes of relaxation are chosen for comparison. This is 
because major changes in the stress state occur during this time interval. Furthermore, this allows us to consider the effect of previous history where the powders with larger $\tr$,
previously could relax longer.

\begin{figure}[!ht]
 \centering
\includegraphics[scale=0.50, angle=270]{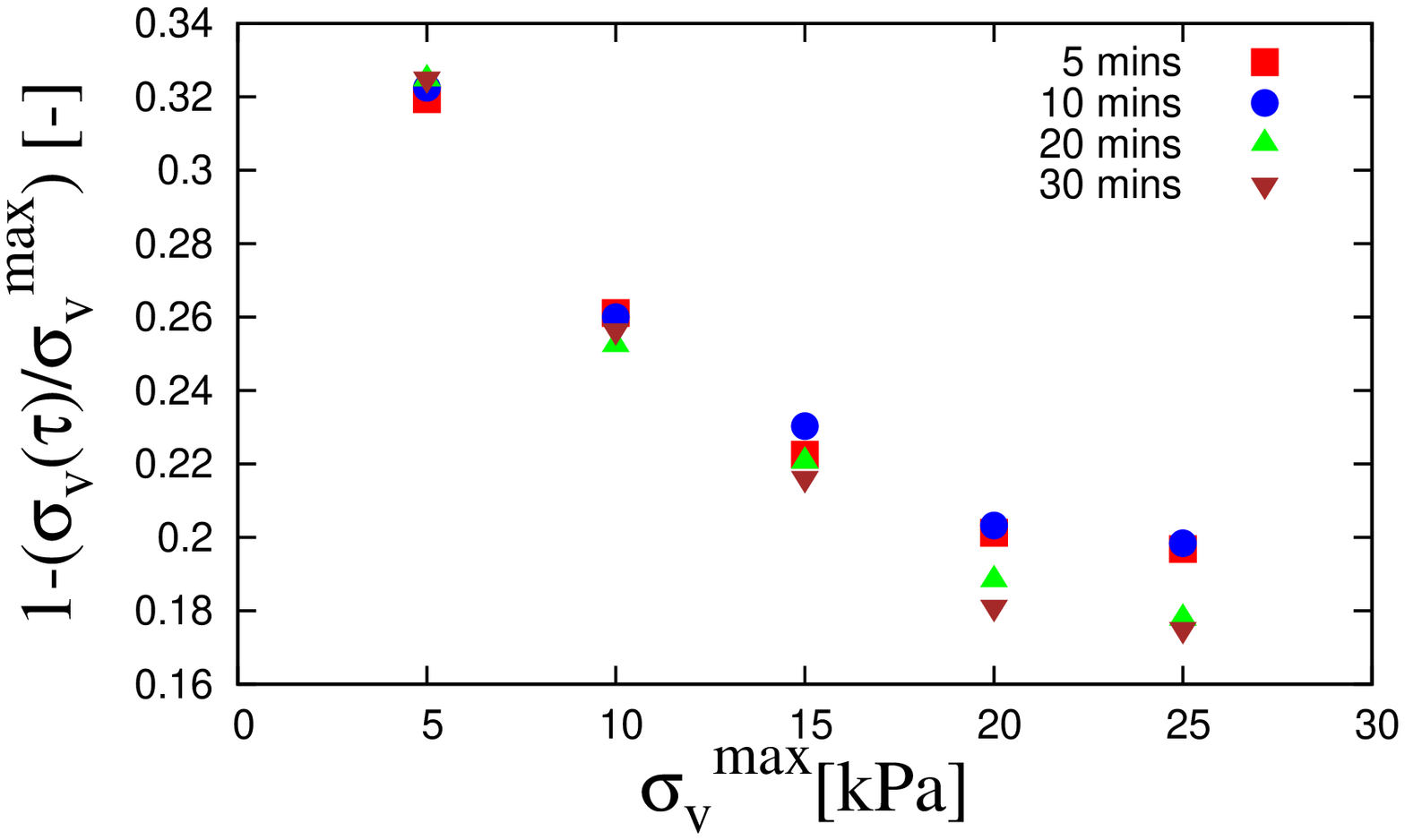}
\caption{Comparison of the relaxed stress states after different relaxation duration $\tr$, as given in the inset. 
Here, only the relative change of stress within the first 5 minutes of the relaxation period is considered.}
\label{compare_relax_time}
\end{figure}

In Fig.\ \ref{compare_relax_time}, we plot the relative stress reduction, $1-(\vstress(\tau=5 {\mathrm{mins}})/\svmax)$ as function of the maximum intermediate stresses $\svmax$ for different previous relaxation times $\tr$.
The relaxation time at the lowest stress of 5 kPa has no effect on the stress reached, as evidenced by the collapse
of the data at $1-(\vstress(\tau)/ \svmax) \approx 0.32$, 
since there is no effect yet of the loading history (different $\tr$); this rather confirms the repeatability of our measurements.

For subsequent relaxations at  higher stresses (10, 15, 20 and 25 kPa), the difference due to the longer previous waiting times becomes visible. 
Consistently, at all stresses, 
an increase in relaxation time $\tr$ results in a lower relative stress reduction. The effects of previously experienced longer relaxation kick 
in at higher stresses, i.e., longer previous loading reduces the possible relaxation in the present state.

In summary, the effect of the relaxation duration becomes visible after the first intermediate stress, when history effects from preceding relaxation stages manifest. At $\svmax = $ 20 and 25 kPa,
the difference between the $\tr=$20 and 30 mins at the highest stress is small suggesting a saturation effect, however, this requires further studies that go beyond the scope of this paper.

\subsection{Dependence on Loading Rate}
\label{sec:strrate}

\begin{figure}[!ht]
 \centering
\subfigure[]{\includegraphics[scale=0.38,angle=270]{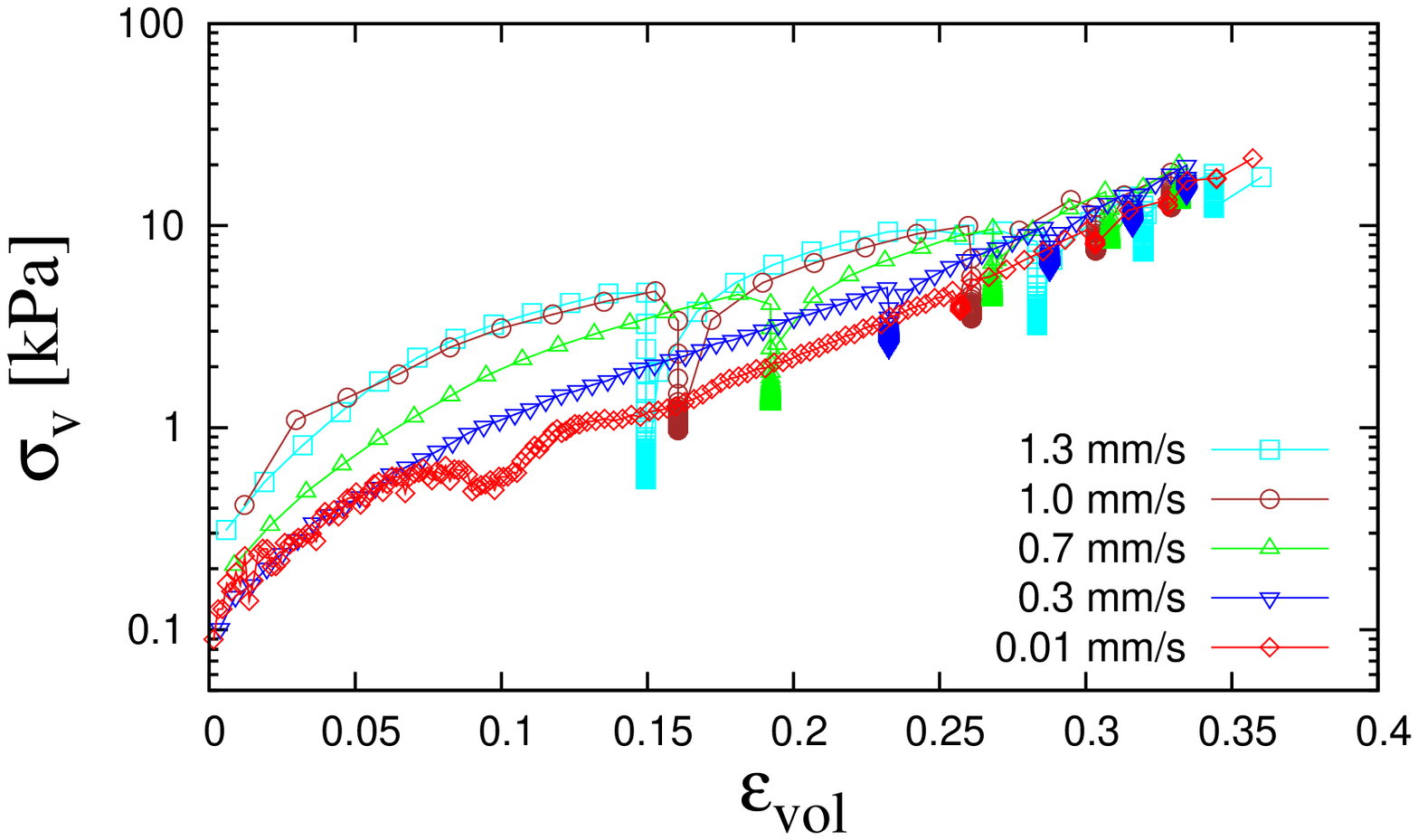}\label{strainrate}}
\subfigure[]{\includegraphics[scale=0.38,angle=270]{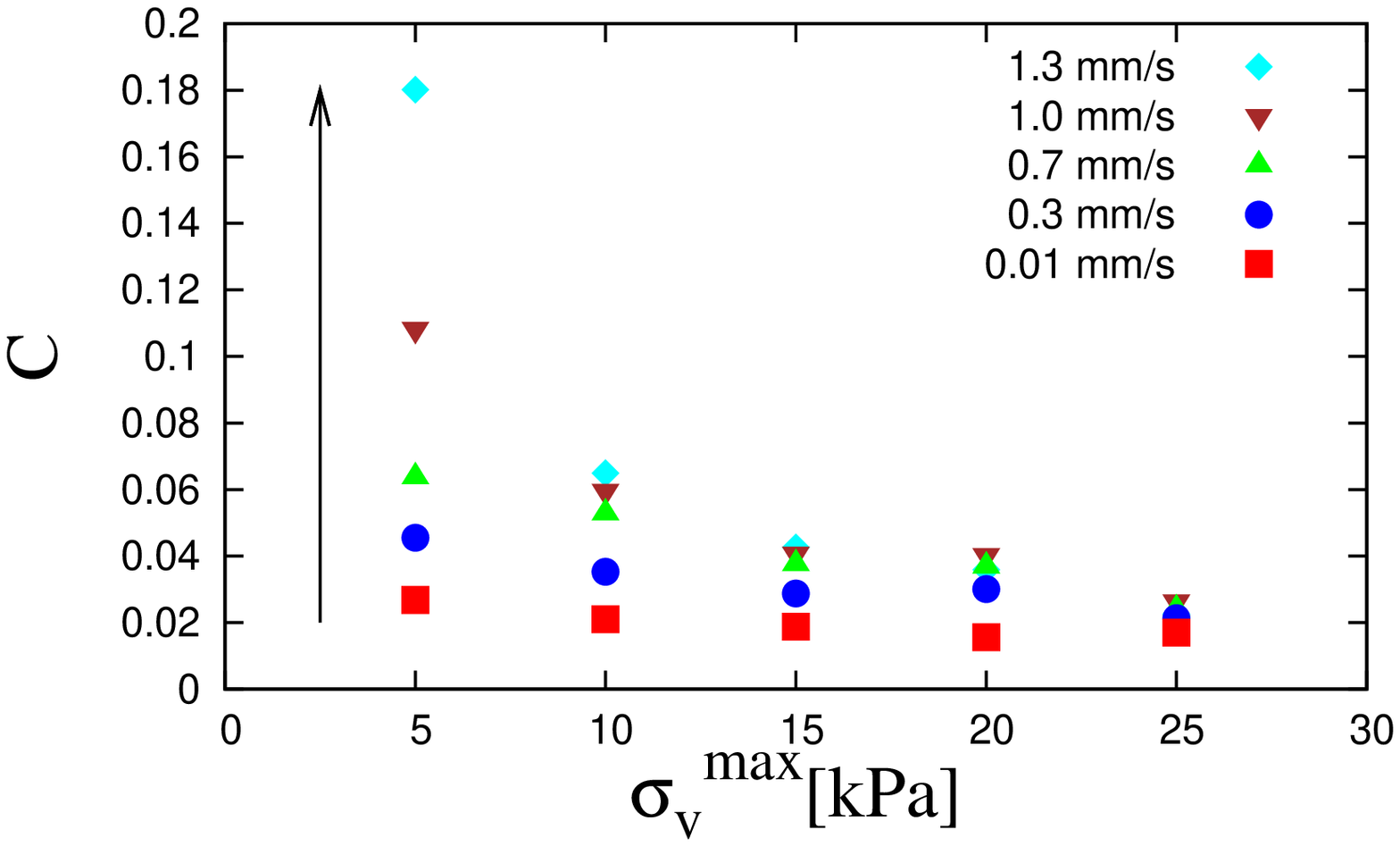}\label{str}}
\caption{ (a) Vertical stress plotted against strain for different loading rates on a semi-log scale. (b) Evolution of the dimensionless relaxation 
parameter $C$ of Eq.\ (\ref{eq:relaxmodel}) with intermediate maximum stresses $\svmax$ for 
different loading rates, as given in the inset. The arrow shows the direction of increasing $C$ with increasing loading rate at each stress level. }
\label{compare_rates}
\end{figure}

In order to investigate the effect of loading rates on the compression and stress relaxation evolution, we use the lambdameter and the limestone sample 
to vary the loading/compression rates. For these experiments, loading rates between 0.01 mm/s and 1.3 mm/s (protocols 5--9 in Table\ \ref{protocoltable}) were studied while 
the relaxation time $\tr$ was set to 10 minutes.
 In Fig.\ \ref{strainrate}, the vertical stress is plotted as function of strain for different loading rates on a semi-logarithmic axis. To allow for clarity,  we show only points at intervals of $\approx$ 20 Hz. 
For small strain ($\evol <$ 0.05), 
we observe a higher stress for increasing loading rate. The strains needed to reach the first stress level $\svmax = 5$ kPa increase with decreasing rate. 
However, all loading rates approach the same strain at the highest stress. This suggests that once the relaxation process is fully underway, 
and the air entrapped in the pores of the particles is released, they approach identical state regardless of the loading rate employed.

Next, using Eq.\ (\ref{eq:relaxmodel}), we fit the different relaxation steps for the different loading rates and present the parameters in Table\ \ref{parametertablec}. The parameter $t_0$
fluctuates and is especially very small in magnitude. 
The values of the dimensionless parameter $C$ for different loading rates and different stress levels, 
listed in Table\ \ref{parametertablec}, are plotted in Fig.\ \ref{str}. For all loading rates, $C$ decreases with increasing stress -- consistent with the findings 
discussed in sections\ \ref{sec:compare_equip} and \ref{sec:powder_comparison}.
At a given stress level, $C$ is found to increase with increasing loading rate as indicated by the arrow. The increase in $C$ with increasing loading rate can be attributed to faster compression
which allows for insufficient relaxation before the next compression stage is initiated. However, the relative increase in $C$ with increasing loading rate is found to decrease with increasing stress. At the highest 
stress (25 kPa), all $C$ values almost collapse on each other indicating an almost identical final relaxed state.

\begin{table}
\caption{Fit parameters for the analytical predictions of the relaxation model Eq.\ (\ref{eq:relaxmodel}) for different loading rates, as presented in Fig.\ \ref{compare_rates}. R1--R5 are the relaxation steps.}
\centering
 \begin{tabular}{llrrr}
    \hline
    Rate&Step & $\svmax$ [kPa] &$t_{0}$ [ms] & $C$ \\ [0.4ex]
    \hline 
   0.01 mm/s &R1 & 4.9960 & 0.0187  & 0.0267 \\ [0.4ex]
    &R2 & 10.0173 & 0.0160 &  0.0209 \\ [0.4ex]
    &R3 & 14.9447 &  0.0285 & 0.0187 \\ [0.4ex]
    &R4 & 19.5277 & 0.0575  & 0.0155  \\ [0.4ex]
    &R5 & 24.9835 & 0.0458 & 0.0170 \\ [0.4ex]
  \hline
   0.3 mm/s &R1 & 4.4068 & 0.0016 & 0.0455 \\ [0.4ex]
    &R2 & 10.001 & 0.00027 &  0.0352 \\ [0.4ex]
    &R3 & 15.0434 &  0.00029 & 0.0287 \\ [0.4ex]
    &R4 & 19.9958 & 0.0042  & 0.0300  \\ [0.4ex]
    &R5 & 22.5748 & 0.00356  & 0.0213 \\ [0.4ex]
  \hline
 0.7 mm/s & R1 & 5.01431 & 0.00002 & 0.0636\\ [0.4ex]
    &R2 & 9.97202 & 0.00041  &  0.0528\\ [0.4ex]
    &R3 & 15.0037 &  0.00043 & 0.0376 \\ [0.4ex]
    &R4 & 20.0097 &  0.00163  & 0.0368  \\ [0.4ex]
    &R5 & 24.5369 & 0.00013  & 0.0239 \\ [0.4ex]
  \hline
   1.0 mm/s &R1 & 5.01512 & 0.00093 &  0.1082  \\ [0.4ex]
    &R2 & 10.0058 & 0.000079 & 0.0596 \\ [0.4ex]
    &R3 & 14.9226 &  0.00023 &  0.0407  \\ [0.4ex]
    &R4 & 20.0208 & 0.0015   & 0.0402 \\ [0.4ex]
    &R5 & 24.1282 & 0.00057  & 0.0263 \\ [0.4ex]
  \hline
   1.3 mm/s &R1 & 4.65097 & 0.0025  & 0.1801 \\ [0.4ex]
    &R2 & 10.10107 & 0.00083 & 0.0648 \\ [0.4ex]
    &R3 & 15.02 &  0.00027 &  0.0425 \\ [0.4ex]
    &R4 & 18.7748 & 0.0020 & 0.0359  \\ [0.4ex]
    &R5 & 24.2288 & 0.00077  & 0.0225 \\ [0.4ex]
  \hline \hline

  \end{tabular}
  \label{parametertablec}

\end{table}

In summary, we find that faster loading rates lead to insufficient time for relaxation with differences most visible at lower stress levels. The effect of loading rate diminishes at higher stress levels. 

\section{Conclusion and Outlook}
\label{sec:creepconclusn}

We have performed oedometric experiments to study the slow relaxation of two cohesive powders under different consolidation stresses. 
One goal was to study the slow relaxation behavior in two experimental devices, namely the custom-built lambdameter and the 
commercially available FT4 Powder Rheometer. Additionally, a comparison of the relaxation behavior of two industrially relevant cohesive powders, namely
cocoa powder with 12\% fat content and Eskal 500 limestone powder, was carried out.


The relaxation behavior i.e., the  decrease in stress occurring at constant volume, is qualitatively reproduced in the two testing equipments. 
On the dependence on aspect ratio, larger strain is required in the setup with higher aspect ratio ($\alpha =$ 1.0) to reach the same intermediate stress in comparison to the setup with lower aspect ratio ($\alpha =$ 0.4).
The relaxation model, cf. Eq.\ (\ref{eq:relaxmodel}), captures perfectly the decrease in stress during relaxation at different
stress levels for both aspect ratios with the response time $t_0$ fluctuating and the dimensionless material parameter $C$ identical for both aspect ratios and systematically decreases from low to high stress levels. 

For the two cohesive powders studied, it is interesting that both materials show an identical response to axial loading  until $\approx$15 percent strain where the difference in the 
response begins to manifest. Eskal 500 limestone is found to produce a softer response to 
applied vertical stress in comparison to cocoa powder, which is probably due to the difference in particle size. At the same stress level, 
cocoa powder is found to relax more slowly but with a larger relative amplitude than Eskal. In terms of the parameters of the model, the response timescale for Eskal, $t_{0E}$, is several orders of magnitude smaller than that of cocoa.
On the other hand, the dimensionless parameter $C$ shows a decreasing trend for both materials and is only about a factor two higher for cocoa than for limestone. 

In terms of the relaxation duration, we find that longer previous relaxation 
leads to observable differences in relative stress reduction $\vstress(\tau = 5 {\mathrm{mins}})/\svmax$, 
reducing the present relaxation. 
Faster loading rates allow for insufficient time for relaxation with differences in the dimensionless parameter $C$ most visible for relaxation at low stresses.
The effects of loading rate are attenuated as stress is increased. 

Further studies will focus on the comparison between the two testing devices for identical aspect ratios and the solution of the model for finite compression rate. The effects of system walls  of 
the experimental devices also needs to be given further attention. The validity of the proposed model for relaxation at constant stress (or strain creep) will be investigated. Finally, the incorporation of the features of 
the present findings into discrete element contact models for cohesive powders will be explored.


\section{Appendix A: Testing Equipments - A Comparison}
\label{sec:compare_equip}

A comparison of testers is necessary for several reasons. Apart from the fact that several literatures have reported on comparative studies between 
different testers used in the characterization of cohesive powders \cite{janssen2003measurements,kamath1994flow, thakur2013characterization}, most differences observed have been attributable to human errors,
differences in the filling procedure and the measurement conditions. For our experiments, it is important to confirm that the relaxation feature can be
reproduced in different testers and is not due to drift or bias in our testing equipments. The material used for this comparison is cocoa powder.

\begin{figure}[!ht]
 \centering
\subfigure[]{\includegraphics[scale=0.37,angle=270]{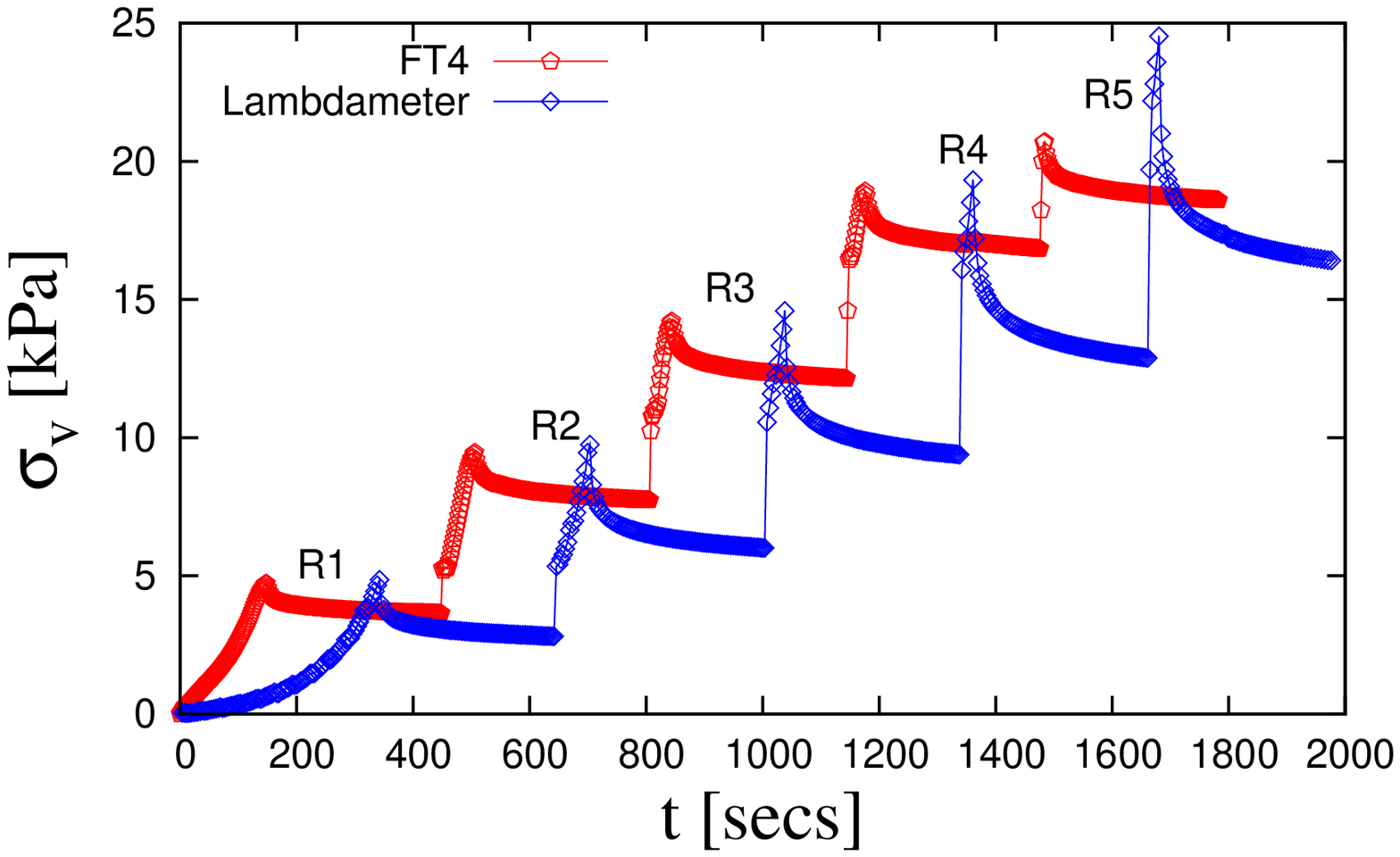}\label{comp_cocoa_time}}
\subfigure[]{\includegraphics[scale=0.37,angle=270]{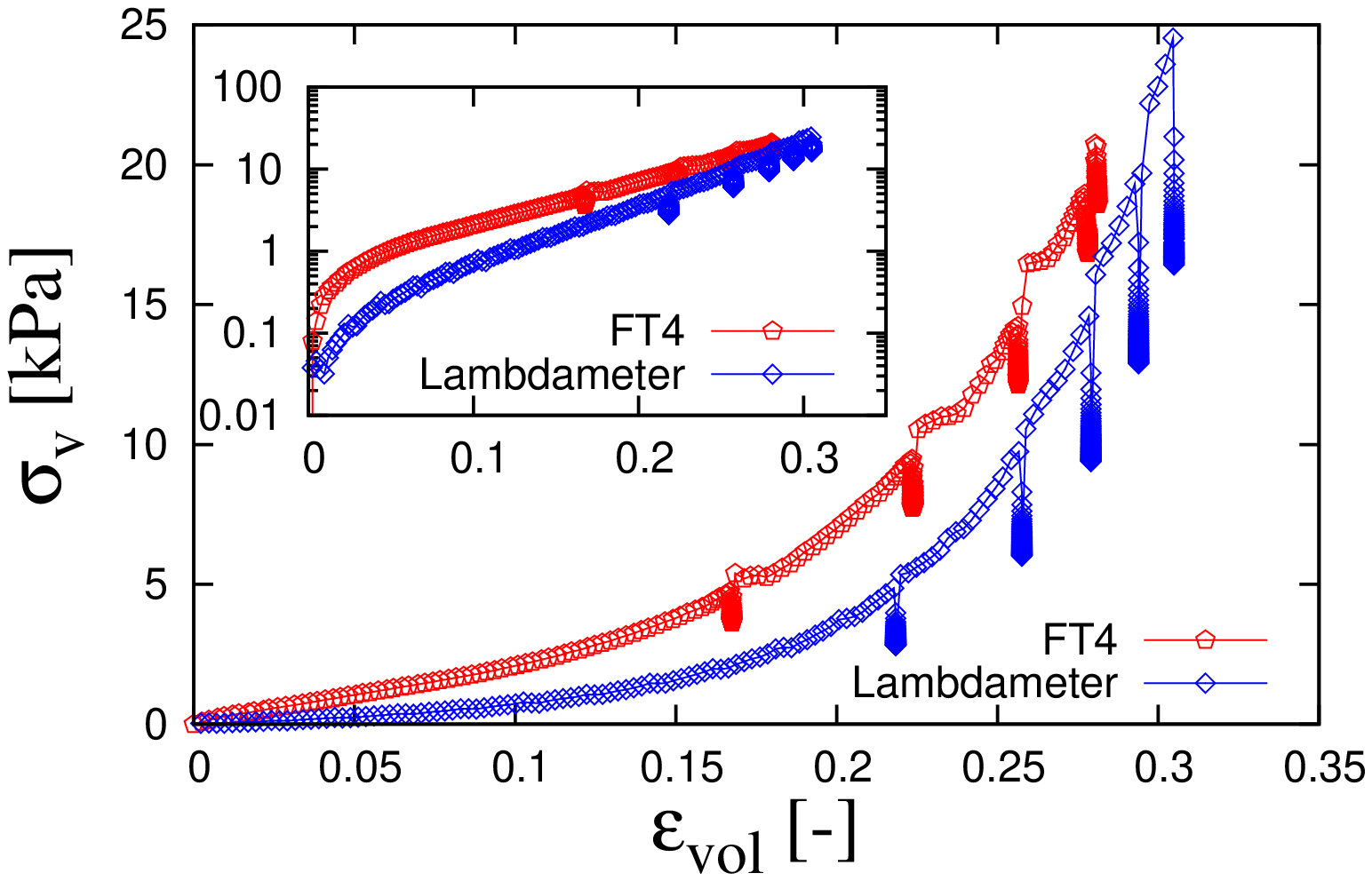}\label{comp_cocoa_strain}}
\caption{Comparison of the vertical (axial) stress plotted against (a) time (b) volumetric strain for experiments with cocoa powder using the FT4 Powder Rheometer and the lambdameter.
The carriage velocity is 0.05 mm/s while R1--R5 represent the intermediate relaxations for increasing target stress.}
\label{compare_ft4_lambda}
\end{figure}

In order to compare the response of the two testing equipments to vertical (axial) stress, we perform uniaxial compression test on cocoa powder
with a carriage speed of 0.05mm/s (protocol 1 in Table \ref{protocoltable}). 
Five intermediate relaxation stages (R1--R5){\footnote[1]{Due to the difference in the final stress reached at 
R5 in the FT4 (22 kPa) and lambdameter (25 kPa), the final relaxation R5 for both equipments should be compared with caution.}}, 
in which the top piston/punch is held in position 
for 300 seconds at specific intervals of 5 kPa during the compression tests are included. In Fig.\ \ref{comp_cocoa_time} we plot
the vertical stress $\vstress$ as function of time. During loading, the axial stress builds up with time until the first target stress of 5 kPa (at R1) is reached.
We observe a slower increase of the axial stress in the lambdameter in comparison to the FT4 Powder Rheometer even though the respective pistons were moved with the same 
speed. Consequently, the vertical compaction and thus strain at constant vertical stress is higher in the lambdameter than in the FT4 Powder Rheometer. 
This is possibly due to the difference in aspect ratio of the experimental moulds and filling procedures (e.g.\ conditioning of the sample by a rotating blade in case of the FT4 Powder Rheometer) resulting in different masses for both equipments, 
leading to different initial densities of the same sample, and consequently producing different response to compressive stress.  


With the initiation of the first relaxation R1 at 5 kPa, we observe for both equipments a time-dependent stress relaxation during the rest-time of 300 seconds. 
Moreover, according to the equation of Janssen \cite{janssen1895versuche}, due to the larger diameter of the lambdameter, the stress away from the powder surface is larger, 
resulting in higher mean vertical stresses. 
This observation,
along with other observations reported in literature for other granular materials \cite{schulze2003time, zetzener2002relaxation}, confirms that the stress relaxation is not due to a drift in the measuring equipments but it is 
a material feature as discussed in section\ \ref{sec:compare_equip_aspectratio}. 
From 5 kPa, we observe an approximate 45 percent relative decrease in stress for the lambdameter compared to 22 percent in the FT4 Powder Rheometer. 
Due to the non-porous lid
and the larger diameter of the lambdameter, at similar height, the escape of the air trapped and compressed in the powder takes more time.
With the activation
of axial compression after the relaxation, we observe a sharp increase in the axial stress until the next intermediate stress state is reached.

The evolution of stress and strain in both testing equipments is shown in  Fig.\ \ref{comp_cocoa_strain}, where the vertical stress is plotted against volumetric strain. 
We observe that the lambdameter initially produces a softer response to the applied stress, as evidenced by the slower increase in the vertical stress during loading. 
At higher intermediate strain, similar stress increase with strain is observed in the lambdameter as compared to the FT4 Powder Rheometer. The comparison of the response
of both testing equipments for identical aspect ratio is a subject for future work and will be presented elsewhere.

\section*{Acknowledgement}
Helpful discussions with N. Kumar and M. Wojtkowski are gratefully appreciated.
This work is financially supported by the European Union funded 
Marie Curie Initial Training Network, FP7 (ITN-238577),
see {http://www.pardem.eu/} for more information.




%
%
%






%



\end{document}